\documentclass[%
reprint,
superscriptaddress,
amsmath,amssymb,
aps,
]{revtex4-2}

\usepackage{graphicx}
\usepackage{dcolumn}
\usepackage{braket}
\usepackage{bm}
\usepackage{hyperref}

\usepackage{xcolor}

\usepackage{cleveref}
\usepackage{multirow}

\begin{document}
	
	\preprint{APS/123-QED}
	
	\title{Analog Circuit-QED Simulator of Quantum Spin Dynamics Through the Extended Bose-Hubbard Model}
	
	\author{Ivan V. Dudinets}
	\email{dudinets@phystech.edu}
	\affiliation{Russian Quantum Center, Skolkovo, Moscow 121205, Russia}
	\affiliation{Moscow Institute of Physics and Technology, Institutskii per. 9, Dolgoprudnyi, 141700, Russia}
	
	\author{Jaehee Kim}%
	\affiliation{SKKU Advanced Institute of Nanotechnology (SAINT), Sungkyunkwan University, Suwon 16419, Republic of Korea}%
	
	\author{Tom\'as Ramos}
	\affiliation{Institute of Fundamental Physics IFF-CSIC, Calle Serrano 113b, 28006 Madrid, Spain}%
	
	\author{Aleksey K. Fedorov}
	\affiliation{Russian Quantum Center, Skolkovo, Moscow 121205, Russia}
	\affiliation{National University of Science and Technology ``MISIS”, Moscow 119049, Russia}
	\affiliation{Lebedev Physical Institute, Russian Academy of Sciences,
		Leninskii Prospect 53, Moscow 119991, Russia}
	
	\author{Vladimir I. Man'ko}
	\affiliation{Lebedev Physical Institute, Russian Academy of Sciences,
		Leninskii Prospect 53, Moscow 119991, Russia}
    \affiliation{Moscow Institute of Physics and Technology, Institutskii per. 9, Dolgoprudnyi, 141700, Russia}
	
\author{Joonsuk Huh}
\email{joonsukhuh@yonsei.ac.kr}
\affiliation{Department of Chemistry, Yonsei University, Seoul 03722, Republic of Korea}
\affiliation{Department of Quantum Information, Yonsei University, Incheon 21983, Republic of Korea}

	\date{\today}
	
	\begin{abstract}
        We propose and validate a framework for analog simulation of the Heisenberg spin model using 
a circuit quantum electrodynamics (circuit-QED) platform.  To this end, we develop a continuous family of deformed boson representations of the 
SU(2) algebra, which includes the Holstein-Primakoff and Dyson-Maleev transformations as special cases.
 For spin-1/2 systems, we introduce a procedure to circumvent the inherent 
 non-Hermiticity of the representation, showing that this entire family yields the extended Bose-Hubbard (EBH) Hamiltonian. 
 For the experimental realization of this EBH model, we design a scalable circuit-QED architecture based on an engineered Josephson junction array. 
 Numerical simulations confirm that the microwave photon dynamics in this simulator accurately reproduces the original spin dynamics. Our work establishes an
experimentally accessible method for investigating complex quantum spin dynamics in a highly controllable bosonic setting.
	\end{abstract}
	
	\maketitle
	
	
	\section{Introduction}
	
	The dynamics of quantum spin systems lie at the heart of various pivotal phenomena in condensed matter physics~\cite{chaikin1995principles}, quantum information science~\cite{Feynman1982,bennett2000quantum}, 
	and materials science~\cite{felser2007spintronics,wolf2001spintronics}. 
	Many-body spin models, such as the Heisenberg spin model~\cite{heisenberg1985theorie}, play a fundamental role in understanding magnetic ordering~\cite{Thomson1992}, phase transitions~\cite{Sachdev2011}, 
	and quantum many-body effects~\cite{newell1953theory,kanamori1966magnetization,auerbach2012interacting,pires2021theoretical}. 
	These models also have the potential to explain intriguing phenomena, such as high-temperature superconductivity~\cite{emery1988mechanism} and quantum criticality~\cite{Sachdev2011,sachdev2011quantum, sachdev2008quantum}. 
    
	However, direct simulation of spin dynamics in the high-dimensional Hilbert spaces of these models poses significant computational challenges~\cite{park2023hardness}, 
	which often exceed the capabilities of classical supercomputers~\cite{Rosner2003,troyer2005computational} and noisy intermediate-scale gate-based quantum processors~\cite{Zhukov2018,Brysen2025,Yamanouchi2025}. We note that the latter case requires not only a sufficient number of qubits but also high-fidelity gate operations, because the number of operations 
(steps in the Trotter decomposition) in such a setting is significant~\cite{Zhukov2018}.
	Overcoming these difficulties may require quantum error correction~\cite{Martinis2012,Fedorov2022}.
	
	A promising approach to address these challenges is to use analog quantum simulation~\cite{buluta2009quantum,daley2022practical}.
	The idea is to emulate a complex quantum system, e.g., solid matter, by another quantum system with well-known and controllable properties, such as an ensemble of cold atoms trapped in an optical field~\cite{daley2022practical}.
    A particularly versatile platform for this purpose is the Bose-Hubbard (BH) model~\cite{hubbard1963electron}, which describes interacting bosons on a lattice and has been used to simulate quantum systems ranging from condensed matter models~\cite{Bloch2002,essler2005one,Bloch2010} to lattice gauge theories~\cite{aidelsburger2022cold}. 

    Quantum simulation has been successfully realized in various physical setups, 
including ultracold atoms in optical lattices~\cite{Bloch2008,Bloch2010,Bloch2012,Lukin2017,Demler2019,Browaeys2022,Luo2025} 
and trapped ions~\cite{Blatt2012,Monroe2017}. 
Among the most promising emerging platforms are circuit quantum electrodynamics (QED) 
systems~\cite{Houck2012,Martinis2015,Hartmann2016,blais2021circuit, 10.1093/nsr/nwaf246, PRXQuantum.2.040202,PhysRevApplied.9.064029}. Their advanced architectures and readout techniques, 
along with precise control over Hamiltonian parameters, make them an ideal platform 
for analog quantum simulation, as demonstrated in various studies  \cite{PhysRevA.95.042313,PhysRevB.99.134308,blais2021circuit,PRXQuantum.3.040314,PhysRevB.107.174306}. Specifically, microwave photons in these systems can
 emulate bosonic degrees of freedom, which has enabled the implementation 
 of the BH model~\cite{schmidt2013circuit,yanay2020two,wilkinson2020superconducting}. 
These recent advancements open new opportunities for simulating complex quantum spin 
dynamics of Heisenberg models in an experimentally feasible and scalable manner.
	
In this work, we explore an approach to simulating spin-$1/2$ dynamics of the Heisenberg model using a circuit quantum electrodynamics (circuit-QED) framework. The key idea is to represent spin operators in terms of bosonic operators. To this end, we develop a continuous family of deformed boson representations of the SU(2) algebra, which includes the Holstein--Primakoff (HP)~\cite{holstein1940field} and Dyson--Maleev (DM)~\cite{dyson1956general, dyson1956thermodynamic, maleev1958scattering} transformations as special cases. The construction is based on $f$-deformed bosonic operators~\cite{VIManko_1997, Dudinets_2017} and is continuously parametrized by $\alpha \in (0,1]$, where the classical HP and DM cases are reproduced at $\alpha = 1/2$ and $\alpha = 1$, respectively. Importantly, this family is derived for arbitrary spin $S$, with the physical state space obtained by truncating the infinite-dimensional bosonic Hilbert space to the subspace spanned by $\ket{n}$ ($n = 0,1,\ldots, 2S$).

We then apply this framework to the Heisenberg model for spin-$1/2$ systems and demonstrate that the entire family yields the same extended Bose-Hubbard (EBH) Hamiltonian. While the mapping is generally non-Hermitian for $\alpha \neq 1/2$, we propose a method to render it Hermitian. This crucial step ensures that the resulting bosonic system accurately reproduces the quantum dynamics of the original spin system, thereby enabling its analog simulation within a circuit-QED framework.

    We validate the bosonic representation through numerical simulations of a circuit-QED system acting as a bosonic simulator. Our results demonstrate that the microwave photon dynamics in this system accurately reproduces the spin dynamics of the original Heisenberg model.

    Building on this validated mapping, we propose a scalable circuit-QED architecture to physically realize the EBH model. Our design is based on an engineered array of Josephson junctions \cite{RamosJJA} --- a platform initially explored for directional quantum amplifiers \cite{ramos_topological_2021,GomezLeon2022}. We establish a complete framework for simulating the Heisenberg model by means of its EBH representation on a circuit-QED platform. Our approach bridges the rich physics of quantum spin systems with 
the experimental capabilities of circuit-QED, offering a practical platform for future exploration of complex quantum phenomena.

	The paper is structured as follows. 
	In Sec.~\ref{sec:map}, we derive the bosonic (EBH) Hamiltonian for the Heisenberg model using the DM transformation.
	In Sec.~\ref{sec:JJA_system}, we present the design of a circuit-QED device capable of simulating this EBH model. In Sec.~\ref{sec:correspondence}, we characterize the parameters of the proposed device. 
	In Sec.~\ref{sec:numerics}, we validate the mapping by performing numerical simulations. 
    Finally, we summarize and discuss our results in Sec.~\ref{sec:discussion}.

    \section{methods}
\subsection{Deformed boson representation of spin operators\label{sec:boson_repr}}

The spin operators $\hat{S}^+$, $\hat{S}^-$ and $\hat{S}^z$ satisfy the commutation relations of the $SU(2)$ algebra:

\begin{align}
    [\hat{S}^z, \hat{S}^{\pm}] &= \pm \hat{S}^{\pm}, \label{eq:commutators_sz} \\
    [\hat{S}^+, \hat{S}^-] &= 2\hat{S}^z. \label{eq:commutators_spsm}
\end{align}

We construct a representation of the spin operators in terms of bosonic creation $\hat{a}^{\dagger}$ and annihilation $\hat{a}$ operators satisfying $[\hat{a}, \hat{a}^{\dagger}] = 1$. To this end, we introduce the $f$-deformed bosonic operators defined as:
\begin{align}
    \hat{A}_p^\dagger &= \hat{a}^\dagger f_p(\hat{n}), \label{eq:f_osc_plus}\\
    \hat{A}_m &= f_m(\hat{n})\hat{a}, \label{eq:f_osc_minus} 
\end{align}
where $f_{p}(\hat{n})$ and $f_{m}(\hat{n})$ are deformation functions depending on the number operator $\hat{n} = \hat{a}^{\dagger}\hat{a}$.

To establish a connection with the spin algebra, we evaluate the commutator between the deformed operators:
\begin{equation}
[\hat{A}_p^\dagger, \hat{A}_m] = \hat{n}\, F(\hat{n}-1) - (\hat{n}+1)\, F(\hat{n}),
\end{equation}
where
$F(\hat{n}) = f_p(\hat{n}) f_m(\hat{n})$.
We impose a condition on the deformed boson algebra by requiring this commutator to be a linear function of $\hat{n}$ of the form  $2(\hat{n} - S)$, where $S$ is the spin magnitude. This requirement leads to the finite-difference equation:
\begin{equation}
    \hat{n} F(\hat{n}-1) - (\hat{n}+1) F(\hat{n}) = 2(\hat{n}-S ),
\end{equation}
which is satisfied by 
    $F(\hat{n}) = 2S- \hat{n}.$
This result allows us to introduce the spin projection operator 
\begin{equation}
    \hat{S}^z = \hat{n} - S.\label{eq:general_sz}
\end{equation}
The commutation relations \eqref{eq:commutators_sz} follow directly from the  identities $f(\hat{n})\hat{a} = \hat{a}f(\hat{n}-1)$ and $f(\hat{n})\hat{a}^{\dagger} = \hat{a}^{\dagger}f(\hat{n}+1)$, which hold for both $f_p$ and $f_m$.
By identifying the remaining spin operators as $\hat{S}^+ = \hat{A}_p^\dagger$ and $\hat{S}^- = \hat{A}_m$, the mapping takes the form:
\begin{align}
    \hat{S}^+ &= \hat{a}^{\dagger}f_p(\hat{n}), \label{eq:general_sp} \\
    \hat{S}^- &= f_m(\hat{n})\,\hat{a},  \label{eq:general_sm}
\end{align}
where the deformation functions are related by
\begin{equation}
    f_p(\hat{n}) f_m(\hat{n}) = 2S - \hat{n}. \label{eq:general_condition}
\end{equation}
The space of spin-$S$ states is finite-dimensional, so the infinite-dimensional Hilbert space of bosons must be truncated. Hereinafter, the space spanned by the eigenstates  $\ket{n}$ ($n=0,1,\ldots, 2S$) of the number operator is referred to as the physical state space.
On this subspace, we have $\hat{S}^+ \ket{2S} = 0$ and $\hat{S}^- \ket{0} = 0$. The first condition yields
\begin{equation}
    f_p(2S) = 0, \label{eq:fp_boundary}
\end{equation}
while the second one holds automatically.
Equations \eqref{eq:general_sz}–\eqref{eq:general_sm}, supplemented by conditions \eqref{eq:general_condition} and \eqref{eq:fp_boundary}, define a broad class of bosonic mappings.
The Holstein--Primakoff (HP) transformation~\cite{holstein1940field} corresponds to the deformation functions
\begin{equation}
f_p(\hat{n}) = f_m(\hat{n}) = \sqrt{2S}\sqrt{1-\frac{\hat{n}}{2S}}, \label{eq:HP_functions}
\end{equation}
while the Dyson--Maleev (DM) transformation~\cite{dyson1956general, dyson1956thermodynamic, maleev1958scattering} is reproduced by
\begin{equation}
f_p(\hat{n}) = \sqrt{2S}\left(1-\frac{\hat{n}}{2S}\right), \qquad f_m(\hat{n}) = \sqrt{2S}. \label{eq:DM_functions}
\end{equation}

As a generalization of these classical transformations, we introduce a continuous family of deformed boson representations parametrized by $\alpha$ ($0 < \alpha \le 1$):
\begin{align}
    \hat{S}^+ &= \sqrt{2S}\,\hat{a}^{\dagger} \left(1 - \frac{\hat{n}}{2S}\right)^{\alpha}, \\
    \hat{S}^- &= \sqrt{2S}\,\left(1 - \frac{\hat{n}}{2S}\right)^{1-\alpha}\hat{a},
\end{align}
with $\hat{S}^z$ being determined by Eq.~\eqref{eq:general_sz}.
This representation reproduces the correct spin commutation relations~\eqref{eq:commutators_sz} and \eqref{eq:commutators_spsm}. For any $\alpha \neq 1/2$, 
the operators $\hat{S}^-$ and $\hat{S}^+$ are not Hermitian conjugates of each other, which leads to a non-Hermitian mapping.
While the two classical cases correspond to $\alpha = 1/2$ and $\alpha = 1$, all other values of $\alpha$ define alternative, non-standard representations.

\subsection{Boson encoding of the Heisenberg model\label{sec:map}}
	We start from the Heisenberg Hamiltonian in an external magnetic field, i.e.,  
	\begin{equation}
		\label{eq:H_spin}
		\hat{H} = - \sum_{\langle j,k\rangle} J_{jk}\left(\hat{S}^x_j\hat{S}^x_k+\hat{S}^y_j\hat{S}^y_k+\hat{S}^z_j\hat{S}^z_k\right) 
		+\sum_{j=1}^N h_j \hat{S}^z_j,
	\end{equation}
	where $\hat{S}^{x}_j$, $\hat{S}^{y}_j$, $\hat{S}^{z}_j$ are the components of spin-$1/2$
    at a lattice site $j$.  
	The coupling constants $J_{jk}$  determine the interaction strength between spins, 
	$h_j$ refer to the magnetic field strength, and the sum is taken over connected sites of the lattice. 
    Although the Heisenberg model is exactly solvable when the spins are arranged in a one-dimensional chain~\cite{YangYang1966}, 
	 simulating the model on classical computers in the case of higher lattice dimensions is computationally hard because of the exponential growth in the number of system states. 
	
	In this study, we translate the Heisenberg Hamiltonian into a form compatible with bosonic devices, focusing on the circuit quantum electrodynamics (circuit-QED) platform. 
	We begin by rewriting the Heisenberg model in the following form:
    \begin{equation}
\label{eq:H_spin_plus}
\begin{split}
    \hat{H} = &-\sum_{\langle j,k\rangle} J_{jk} \left(\frac{1}{2}\hat{S}^+_j\hat{S}^-_k+\frac{1}{2}\hat{S}^-_j\hat{S}^+_k+\hat{S}^z_j\hat{S}^z_k\right) \\
    &+\sum_{j=1}^N h_j \hat{S}^z_j,
\end{split}
\end{equation}
	where we have introduced the raising and lowering spin operators $\hat{S}^{\pm}_j = \hat{S}^x_j \pm i\hat{S}^y_j$ that obey the commutation relations:
\begin{align}
    [\hat{S}^z_j, \hat{S}^{\pm}_k] &= \pm \hat{S}^{\pm}_k \delta_{jk}, \label{eq:commutators_spin_sz} \\
    [\hat{S}^+_j, \hat{S}^{-}_k] &= 2 \hat{S}^{z}_j \delta_{jk}. \label{eq:commutators_spin_spsm}
\end{align}
	
	The simulation of spin systems on a circuit-QED platform necessitates reformulating the system Hamiltonian in terms of bosonic degrees of freedom.  
	In our study, we use the deformed boson transformation, which allows us to represent spin operators through bosonic creation $\hat{a}^{\dagger}_j$ and annihilation $\hat{a}_j$ operators as follows:  
    \begin{align}
    \hat{S}^+_j &= \hat{a}^{\dagger}_j \left(1-\hat{n}_j\right)^{\alpha}, \label{eq:alpha_mapping_plus} \\
    \hat{S}^-_j &= \left(1-\hat{n}_j\right)^{1-\alpha}\hat{a}_j, \label{eq:alpha_mapping_minus} \\
    \hat{S}^z_j &= \hat{n}_j - \frac{1}{2},\label{eq:alpha_mapping_z}
\end{align}
	where $\hat{n}_j = \hat{a}^{\dagger}_j\hat{a}_j$ stands for the number operator at a site $j$. 
    For a multi-site system, the physical state space is spanned by the states of the form
$\ket{n_1,n_2,\ldots,n_N}$ with $n_j=0,1$ for all $j$. Here, $\ket{0_j}$ and $\ket{1_j}$ are the eigenstates of the number operator $\hat{n}_j$. The truncation of the Hilbert space is guaranteed by
preparing 
the initial state within the physical subspace
and by ensuring that the quantum dynamics maintains the system within this subspace (see Sec. \ref{sec:discussion}). On this subspace, the mapping reduces to
    \begin{align}
    \label{eq:HP}
    \hat{S}^+_j &= \hat{a}^{\dagger}_j\left(1-\hat{n}_j\right), \\
    \hat{S}^-_j &= \left(1-\hat{n}_j\right)\hat{a}_j,
\end{align}
with $\hat{S}^z_j$ given by Eq.~\eqref{eq:alpha_mapping_z}.

To obtain the Hamiltonian of the Heisenberg model expressed in terms of bosonic operators, we exploit the freedom to modify its operator structure by
using terms that vanish in the physical state space. In particular, for $\alpha=1$, which corresponds to the DM case, we proceed by adding terms of the form $\hat{a}^\dagger_j\hat{n}_k\hat{a}_k+\hat{a}^\dagger_k\hat{n}_j\hat{a}_j$, whereas for other values of $\alpha$
we omit operators such as 
$\hat{a}^\dagger_j\hat{n}_j\hat{n}_k\hat{a}_k +\text{h.c.}$. The resulting Hamiltonian reads
    \begin{equation}
\label{eq:H_EBH}
\begin{split}
    \hat{H}_{\rm EBH} = &-\sum_{\langle j,k\rangle} J_{jk}\left(\hat{a}^{\dagger}_j\hat{a}_k + \hat{a}_j\hat{a}^{\dagger}_k\right)\left(1-\frac{\hat{n}_j+\hat{n}_k}{2}\right) \\
    &- \sum_{\langle j,k\rangle} J_{jk}\left(\hat{n}_j-\frac{1}{2}\right)\left(\hat{n}_k-\frac{1}{2}\right) + \sum_{j=1}^N h_j\left(\hat{n}_j-\frac{1}{2}\right).
\end{split}
\end{equation}
	The obtained Hamiltonian coincides with the extended Bose-Hubbard (EBH) model~\cite{sowinski2012dipolar}, which includes the nearest-neighbor interaction and the occupation-induced one-particle tunneling terms.  
	We note that this model has been studied in experiments with ultracold magnetic atoms~\cite{Ferlaino2016}.

An important feature of the EBH system is that its dynamics is confined to the Hilbert subspace spanned by the physical states. Indeed, we observe that each term of the Hamiltonian, when applied to a physical state, produces either zero or another physical state. Consequently, under the proper initial state condition, 
the EBH model reproduces the dynamics of the Heisenberg spin system.
	
	\subsection{Design of the analog circuit-QED-based simulator \label{sec:JJA_system}}
	\begin{figure*}
	\centering
  \includegraphics[width=0.75\linewidth]{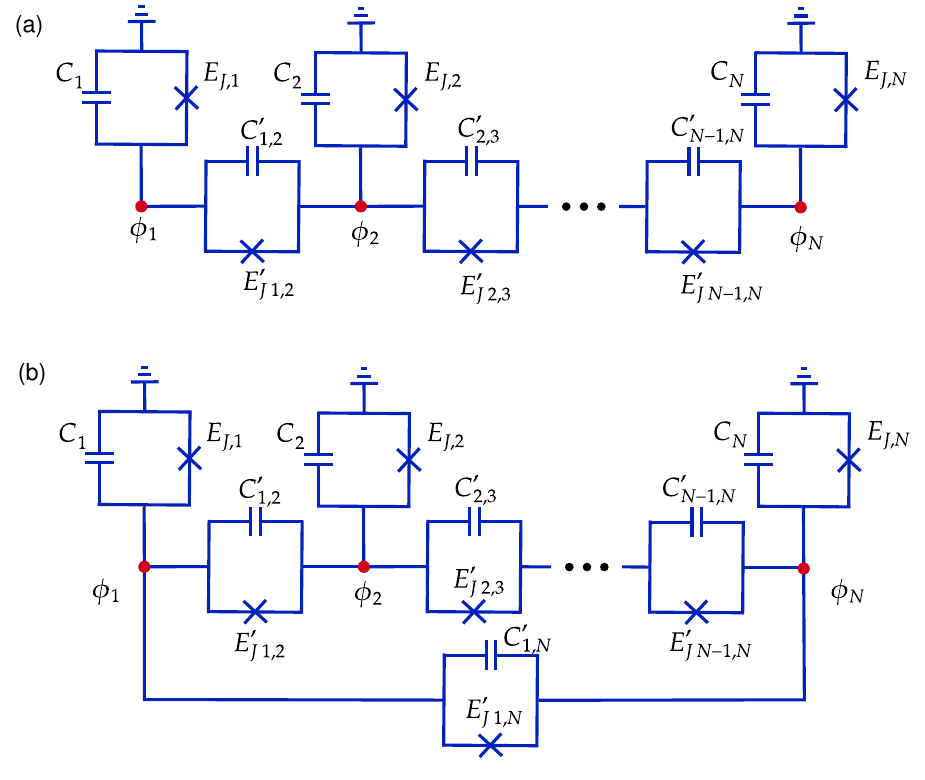}
		\caption{Design of the analog circuit-QED-based simulator for a one-dimensional Heisenberg model with (a) open and (b) periodic boundary conditions. The simulator is the Josephson-junction array that  consists of $N$ Josephson junctions with Josephson energies $E_{J,j}$, each coupled in parallel to a capacitance $C_j$, $j=1,\ldots, N$, and connected to ground. Flux variables $\phi_j$ indicated by red dots play the role of nonlinear microwave oscillators. These nonlinear oscillators interact via capacitive couplings $C'_{jk}$ and nonlinear inductive couplings $E'_{J,jk}$. Adapted from Ref.~\cite{RamosJJA}.
		}
		\label{fig:1D}
\end{figure*}
	\begin{figure*}
		\centering
		\includegraphics[width=0.85\textwidth]{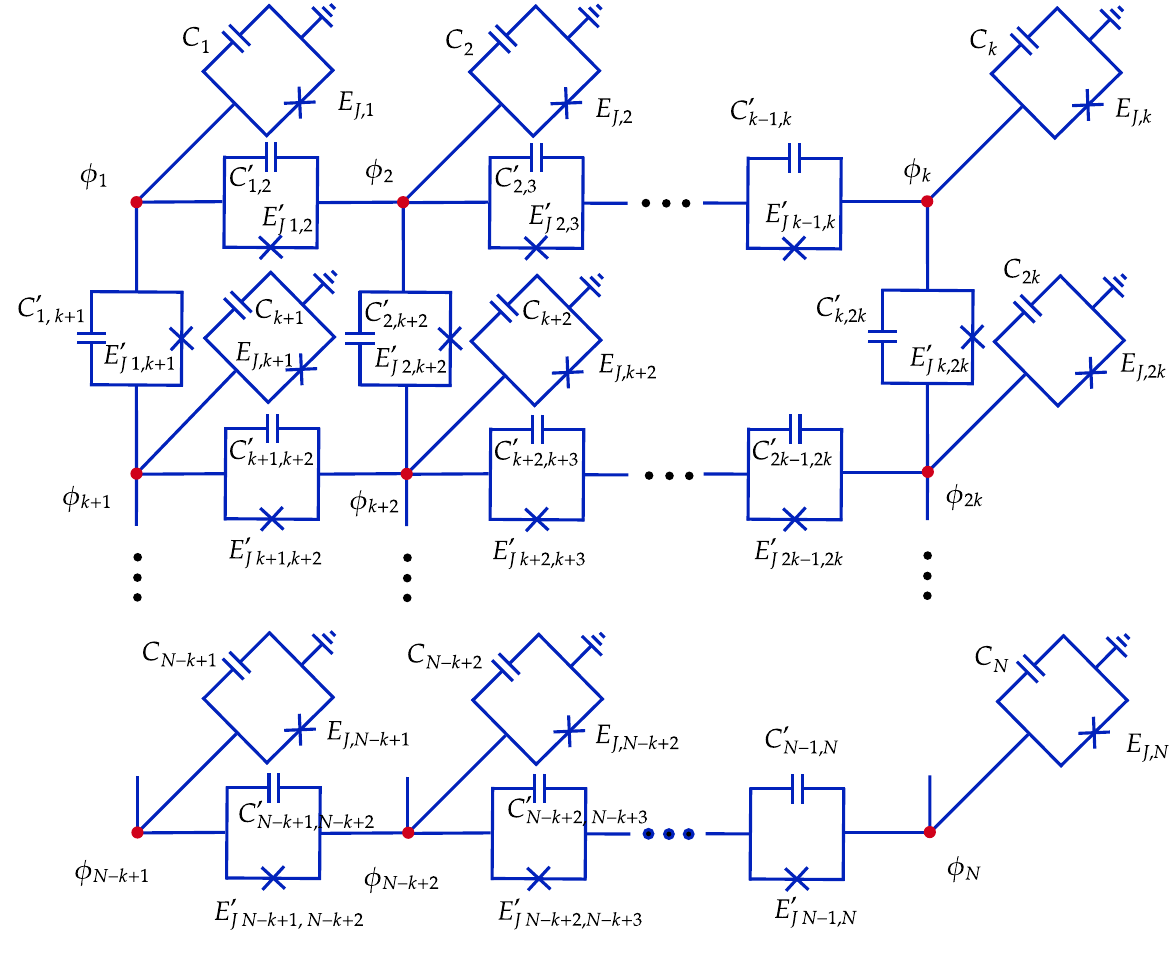}
		\caption{Design of the analog circuit-QED-based simulator for the Heisenberg model of $N$ spins on a rectangular lattice with open boundary conditions. The spins are enumerated row-wise from the top left to the bottom right. Other parameters are the same as in Fig. \ref{fig:1D} 
        }
		\label{fig:2D}
	\end{figure*}
	This section is devoted to the detailed description of the superconducting circuit system that is used as a circuit-QED simulator. The system is a Josephson junction array (JJA) depicted in Fig. \ref{fig:1D} for one-dimensional spin chains with open and 
    periodic boundary conditions, and in Fig. \ref{fig:2D} for a two-dimensional spin lattice. At each site $j$ ($j=1,\ldots, N$) of the array we define a flux variable ${\phi}_j$. This variable represents a nonlinear microwave oscillator formed by a Josephson junction with an energy $E_{J,j}$ in parallel with a capacitance $C_j$, with both elements connected to ground. Each of these nonlinear oscillators interacts with its neighbors by means of a Josephson junction with an energy $E'_{J,jk}$ and a capacitance $C'_{jk}$, $j,k=1,\ldots, N$. The dynamics of the JJA is governed by the Lagrangian \cite{RamosJJA}
	\begin{eqnarray}
		\label{eq:lagrangian}
		{\cal L}_{\rm JJA}&=& {}\sum_{j=1}^N \frac{C_j(\dot{\phi}_j)^2}{2}
		+\sum_{\langle j,k\rangle}\frac{C'_{jk}}{2}(\dot{\phi}_{j}-
		\dot{\phi}_{k})^2\\
		&+&\sum_{j=1}^N E_{J,j}\cos\left(\frac{\phi_j}{\Phi_0}\right)+\sum_{\langle j,k\rangle}E'_{J,jk}\cos\left(\frac{\phi_{j}-\phi_{k}}{\Phi_0}\right),\nonumber
	\end{eqnarray}
	where $\Phi_0 = \hbar/2e$  is the flux quantum with $\hbar$ and $e$ being the reduced Planck’s constant and 
	the electron’s charge, respectively. Double sums are performed over the connected sites of the array. The system Hamiltonian is derived by defining the charge variables $q_j = \partial{\cal L_{\rm JJA}}/\partial\dot{\phi}_j$ and using the Legendre
	transform. In the limit of low coupling capacitances $C'_j \ll C_k$, and the low phase-drop regime $|\phi_j| \ll \Phi_0$, $|\phi_j-\phi_{k}| \ll \Phi_0$, the cosine functions are expanded to give the following approximation for the JJA Hamiltonian:
	\begin{eqnarray}
		\label{eq:H_JJA_approx}
		\hat{H}_{\rm JJA}&=&\sum_{j=1}^N \left(E_{C,j}\,\frac{\hat{q}_j^2}{e^2}
		+E^{\text{eq}}_{J,j}\frac{\hat{\phi}_j^2}{2\,\Phi^2_0}\right) -\sum_{j=1}^NE^{\text{eq}}_{J,j}\frac{\hat{\phi}_j^4}{24\,\Phi^4_0}\nonumber\\
		&+&\sum_{\langle j,k \rangle}\left(E_{\text{coup},jk}\,\frac{\hat{q}_j\,\hat{q}_{k}}{e^2}-E'_{J,jk}\,\frac{\hat{\phi}_j\,\hat{\phi}_{k}}{\Phi^2_0}\right)\\
		&-&\sum_{\langle j,k \rangle} E'_{J,jk}\,\frac{\hat{\phi}_{j}^2\,\hat{\phi}_{k}^2}{4\,\Phi^4_0}
		+\sum_{\langle j,k \rangle} E'_{J,jk}\,\frac{\hat{\phi}_{j}\,\hat{\phi}_{k}}{6\,\Phi^4_0}\left(\hat{\phi}_{j}^2+\hat{\phi}_{k}^2\right).\nonumber
	\end{eqnarray}
	Here, $E^{\text{eq}}_{J,j}=E_{J,j}+\sum_{k\in \text{NN}(j)} E'_{J,jk}$ denotes the equivalent Josephson energy. $\text{NN}(j)$ indicates the sum over the nearest-neighbors of the $j$-th lattice site.
The capacitive energies and the capacitive coupling energies are defined by
	\begin{eqnarray}
		E_{C,j} &=&\frac{e^2}{2\,C^{\text{eq}}_j},\label{eq_appendix:EC}\\
		E_{{\rm coup},jk} &=&\frac{e^2\,C'_{jk}}{C^{\text{eq}}_j\,C^{\text{eq}}_k}\label{eq_appendix:Ecoup},
	\end{eqnarray}
    where $C^{\text{eq}}_j = C_j + \sum_{k\in \text{NN}(j)}C'_{jk}$ is the equivalent capacitance. The system is quantized by promoting  the flux $\phi_j$ and charge $q_j$ variables to commutators satisfying the canonical commutation relation $[q_j, \phi_k]=i\hbar\,\delta_{j, k}$, where $j,k=1,\ldots,N$, and expressing them in terms of the creation and annihilation operators determined by 
	\begin{eqnarray}
		\frac{\hat{\phi}_j}{\Phi_0} &=& \left(\frac{2E_{C,j}}{E^{\rm eq}_{J,j}}\right)^{1/4}\left(\hat{a}_j+\hat{a}^{\dagger}_j\right), \\ 
		\frac{\hat{q}_j}{e} &=& i\left(\frac{E^{\rm eq}_{J,j}}{2E_{C,j}}\right)^{1/4}
		\left(\hat{a}^{\dagger}_j-\hat{a}_j\right).
	\end{eqnarray}
	In the rotating-wave approximation, the quantum JJA Hamiltonian  takes the following form:
	\begin{eqnarray}\label{eq:JJA_Ham}
		\hat{H}_{\rm JJA} &&= \sum_{j=1}^N\left[\left(\hbar\omega_j+\hbar\,\delta\omega_j-\frac{1}{2}\Delta^{\rm eq}_j\right)\hat{n}_j
		+\frac{\hbar\,\delta\omega_j}{2}(\hat{a}^{\dagger}_j)^2\hat{a}^2_j\right]\nonumber\\
		&&+\sum_{\langle j,k\rangle}t_{jk}\left(\hat{a}_{j}^{\dagger}\hat{a}_{k}+\mbox{h.c.}\right)-\sum_{\langle j,k\rangle}\Delta_{jk}\,
		\hat{n}_j\hat{n}_{k}\nonumber\\
		&&+\sum_{\langle j,k\rangle} 
		\left(T_{jk}\,\hat{a}_j\,\hat{n}_j\,\hat{a}^{\dagger}_{k}+\overline{T}_{jk}\,\hat{a}_j\,\hat{n}_{k}\,\hat{a}^{\dagger}_{k}+\mbox{h.c.}\right)\nonumber\\
		&&-\sum_{\langle j,k\rangle}\frac{\Delta_{jk}}{4}
		\left(\hat{a}^2_{k}\left(\hat{a}^{\dagger}_{j}\right)^2+\mbox{h.c.}\right),
	\end{eqnarray}
    where $\Delta^{\rm eq}_{j} =  \sum_{k\in \text{NN}(j)}\Delta_{jk}$ defines the equivalent coupling strength.
	The Hamiltonian describes an array of $N$ nonlinear oscillators with bare frequencies $\omega_j=\sqrt{8E_{C,j}E^{\rm eq}_{J,j}}/\hbar$
	and anharmonicity parameters $\delta\omega_j= -E_{C,j}/\hbar$ and $\Delta^{\rm eq}_j$. 
	The oscillators interact via linear hopping parameters  $t_{jk}$, coupling strengths $\Delta_{jk}$ and nonlinear cross-Kerr couplings $T_{jk}$, $\overline{T}_{jk}$.  These parameters are given by 

    \begin{eqnarray}
t_{jk}&=&E_{\text{coup},jk}\left(\frac{E^{\rm eq}_{J,j}E^{\rm eq}_{J,k}}{4E_{C,j}E_{C,k}}\right)^{\frac{1}{4}}-E'_{J,jk}\left(\frac{4E_{C,j}E_{C,k}}{E^{\rm eq}_{J,j}E^{\rm eq}_{J,k}}\right)^{\frac{1}{4}},
        \\
        \Delta_{jk} &=& 2\,E'_{J,jk}\,
	\left(\frac{E_{C,j}\,E_{C,k}}{E^{\rm eq}_{J,j}\,E^{\rm eq}_{J,k}}\right)^{1/2},
    \\
    T_{jk}&=&E'_{J,jk}\left(\frac{E_{C,j}}{E^{\rm eq}_{J,j}}\right)^{3/4}
	\left(\frac{E_{C,k}}{E^{\rm eq}_{J,k}}\right)^{1/4}\\
    \overline{T}_{jk} &=& E'_{J,jk}\left(\frac{E_{C,j}}{E^{\rm eq}_{J,j}}\right)^{1/4}
	\left(\frac{E_{C,k}}{E^{\rm eq}_{J,k}}\right)^{3/4}.
    \end{eqnarray}
    Appendix~\ref{appendix:H_JJA derivation} provides a detailed derivation of the JJA Hamiltonian.

	\subsection{\label{sec:correspondence}Parameters of the analog circuit-QED-based simulator}
    We demonstrate that the circuit-QED system, described by the JJA Hamiltonian, is equivalent to the Heisenberg model represented in terms of bosonic operators. The equivalence holds 
    for a specific set of parameters of the JJA Hamiltonian. We establish this equivalence by comparing the Hamiltonians of these two systems.

    We restrict the simulation of the circuit-QED system to at most one excitation per site, as this includes all spin$-1/2$ states. This restriction is valid because the initial state condition and the quantum dynamics keep the system within the subspace of physical states  (see Appendix ~\ref{appendix:H_JJA derivation} for details). As a result, the terms with $\hat{a}^2_j$ are absent from the JJA Hamiltonian, namely 
	\begin{eqnarray}
	\label{eq:H_JJA_simplified}
		\hat{H}_{\rm JJA} &&= \sum_{j=1}^N\left(\hbar\omega_j+\hbar\,\delta\omega_j-\frac{1}{2}\Delta^{\rm eq}_j\right)\hat{n}_j
		\\
		&&+\sum_{\langle j,k\rangle}t_{jk}\left(\hat{a}_{j}^{\dagger}\hat{a}_{k}+\mbox{h.c.}\right)-\sum_{\langle j,k\rangle}\Delta_{jk}\,
		\hat{n}_j\hat{n}_{k}\nonumber\\
		&&+\sum_{\langle j,k\rangle} 
		\left(T_{jk}\,\hat{a}_j\,\hat{n}_j\,\hat{a}^{\dagger}_{k}+\overline{T}_{jk}\,\hat{a}_j\,\hat{n}_{k}\,\hat{a}^{\dagger}_{k}+\mbox{h.c.}\right).\nonumber
	\end{eqnarray} In addition, we assume the ratio $E_{C,j}/E^{\rm eq}_{J,j}$ is independent on the index $j$ and is  
	hereafter referred to as $E_{C}/E^{\rm eq}_{J}$. This is realized in practice by having an uniform distribution of the impedance across the lattice (see Appendix~\ref{appendix:parameter_space}). In this case, the parameters of the Hamiltonian are related as follows:
    \begin{eqnarray}
        \Delta_{jk} = 2\,T_{jk}=2\,\overline{T}_{jk}=2\,E'_{J,jk}E_{C}/E^{\rm eq}_{J},
    \end{eqnarray}
     and the linear hopping parameters are given by
	\begin{eqnarray}
	    t_{jk} = E_{\text{coup},jk}\,\left(\frac{E^{\rm eq}_{J}}{2E_{C}}\right)^{1/2}-E'_{J,jk}\,\left(\frac{2E_{C}}{E^{\rm eq}_{J}}\right)^{1/2}.
	\end{eqnarray}
	
	We find that the Hamiltonians of the JJA in Eq.~(\ref{eq:H_JJA_simplified}) and EBH models in Eq. (\ref{eq:H_EBH}) coincide, provided the following correspondence between the parameters is fulfilled:
    
\begin{eqnarray}\label{eq:relation}
		J_{jk} &=& 2\,E'_{J,jk}\,\frac{\,E_{C}}{E^{\rm eq}_{J}},\\
		h_j &=&\left(8E_{C,j}\,E^{\rm eq}_{J,j}\right)^{1/2}-3E_{C,j}+
		\frac{2E_{C}}{E^{\rm eq}_{J}}E_{J,j}.\label{eq:relation2}
\end{eqnarray}
Here, $J_{jk}$ is the coupling constant between spins $j$ and $k$.
	In addition, the parameters of the circuit-QED simulator must satisfy the following condition:   
	\begin{eqnarray}\label{eq:QEDcondition}
		E_{{\rm coup},jk} = 2E'_{J,jk}\frac{E_{C}}{E^{\rm eq}_{J}}
		\left(1-3\left(\frac{E_{C}}{2\,E^{\rm eq}_{J}}\right)^{1/2}\right).
	\end{eqnarray}
In Appendix \ref{appendix:parameter_space}, we analyze the variables of the circuit-QED simulator that satisfy Eqs. ~(\ref{eq:relation}), ~ (\ref{eq:relation2}), ~(\ref{eq:QEDcondition}) and express them in terms of the parameters of the Heisenberg spin model. 

	\section{Numerical Results\label{sec:numerics}}
    We validate our method by comparing the dynamics of the circuit quantum electrodynamics (circuit-QED) simulator described by Eq.~(\ref{eq:H_JJA_simplified}) to those of  various Heisenberg spin models. 
    
    In our simulations, we use the QuSpin package \cite{Weinberg_2017_QuSpin,Weinberg_2019_QuSpin}
to perform exact diagonalization and compute the quantum dynamics of the target spin models and the Josephson junction array (JJA) serving as the simulator. For these spin models, the characteristic interaction strength between spins is on the order of $J = 2\pi\hbar\times 40$ MHz. To realize the JJA model, we use a parameter regime 
centered around typical values for
the capacitive energy $E_C=2\pi\hbar\times 200$ MHz and the Josephson energy $E_J = 2\pi\hbar\times 12.5$ GHz~\cite{blais2021circuit}. The bare frequency of the oscillators is chosen around $\omega = 2\pi\times 5$ GHz. 

In the following subsections, we present several examples of this comparison, showing the dynamics of various observables for different initial states.

\subsection{Magnetization transport in the dimerized antiferromagnetic Heisenberg chain}
We employ the circuit-QED simulator to study the dynamics of a one-dimensional Heisenberg spin chain with different dimerization strengths described by the Hamiltonian
    \begin{equation}
		\label{eq:H1}
		\hat{H}_1 =  J\sum_{j=1}^{N-1}\left(1+(-1)^{j}\delta\right)\left(\hat{S}^x_j\hat{S}^x_{j+1}+\hat{S}^y_j\hat{S}^y_{j+1}+\hat{S}^z_j\hat{S}^z_{j+1}\right),
	\end{equation}
    where the coupling between neighboring spins
    is given by $J(1 + \delta)$ for strong bonds and $J(1 - \delta)$ for weak bonds, with $\delta$ being the dimerization coefficient. 
    This Hamiltonian is of the spin-Peierls type \cite{barford2005peierls, herzog2011dimerized}, which describes 
the interaction between spin degrees of freedom and lattice (phonon) vibrations. Here, $\delta$ is a constant, corresponding to the adiabatic limit of this model, in which phonons are treated classically. We also note that superconducting platforms with dynamical spin-boson coupling (transmon qubits coupled to microwave photons in resonators), relevant to the spin-Peierls model, have been proposed \cite{PhysRevB.89.144508,PhysRevLett.124.190504}.

Initially, the spin system is prepared in the domain-wall state $\ket{\rm DW}_{\text{spin}} = \ket{\uparrow_z\uparrow_z\ldots \uparrow_z\downarrow_z\downarrow_z\ldots\downarrow_z}$, in which spins on the left (right) half are aligned up (down) along the $z$ axis. The time evolution delocalizes the domain wall, thereby leading to magnetization transport. We quantify the spin transport by means of the integrated flow of magnetization through the center \cite{gobert2005real,misguich2017dynamics}
\begin{equation}\label{eq:Mspin}
    \Delta \mathcal{M}_{\text{spin}}(t) = \sum_{j> N/2}^N\left( \langle \hat{S}^z_j\rangle (t)+\frac{1}{2}\right).
\end{equation}
 In addition, we study the spin flip behavior by analyzing the deviations
of the spin expectation values from the initial values 

\begin{eqnarray}\label{eq:mspin}
    \Delta m_{\rm spin}(j,t) = \langle \hat{S}^z_j\rangle (t)-\frac{1}{2}+\theta\left(j-\frac{N}{2}\right),
\end{eqnarray}
where $\theta$ is the Heaviside step function.

In the circuit-QED simulator, the 
domain-wall state is given by $
	\ket{\rm DW}_{\text{boson}} = \ket{1_11_2\ldots1_{N/2}0_{N/2+1}\ldots0_N}$, whereas
the flow of magnetization is represented by the average number of boson excitations in the right part of the JJA 
\begin{equation}\label{eq:Mboson}
    \Delta \mathcal{M}_{\text{boson}}(t) = \sum_{j> N/2}^N\langle \hat{n}_j\rangle (t).
\end{equation}
The local magnetization corresponds to the expectation value of the boson number operators $\Delta m_{\rm boson}(j,t) = \langle \hat{n}_j\rangle (t)+\theta\left(j-N/2\right)$.

\begin{figure*}
		\includegraphics[width=0.45\linewidth]{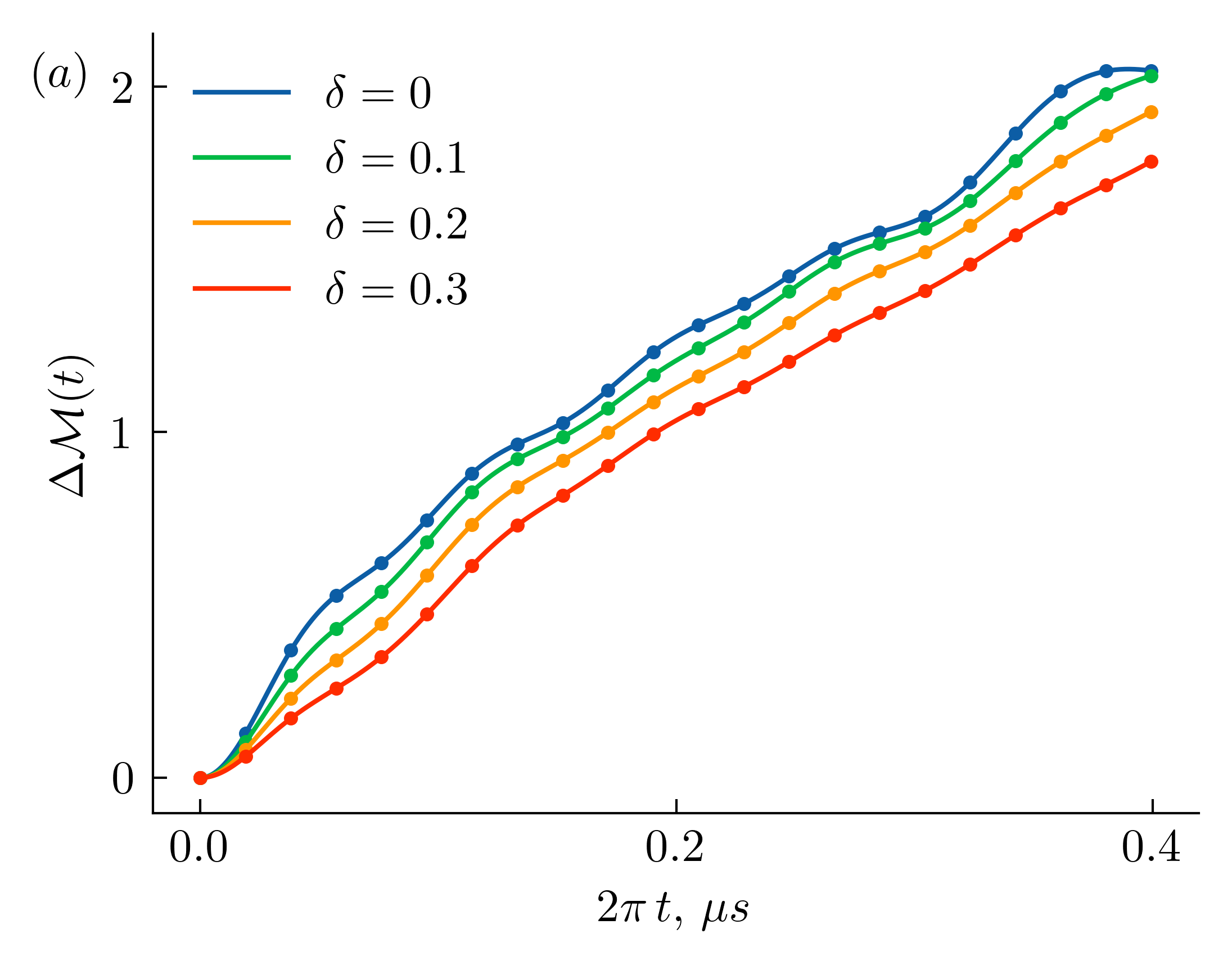}
        \includegraphics[width=0.45\linewidth]{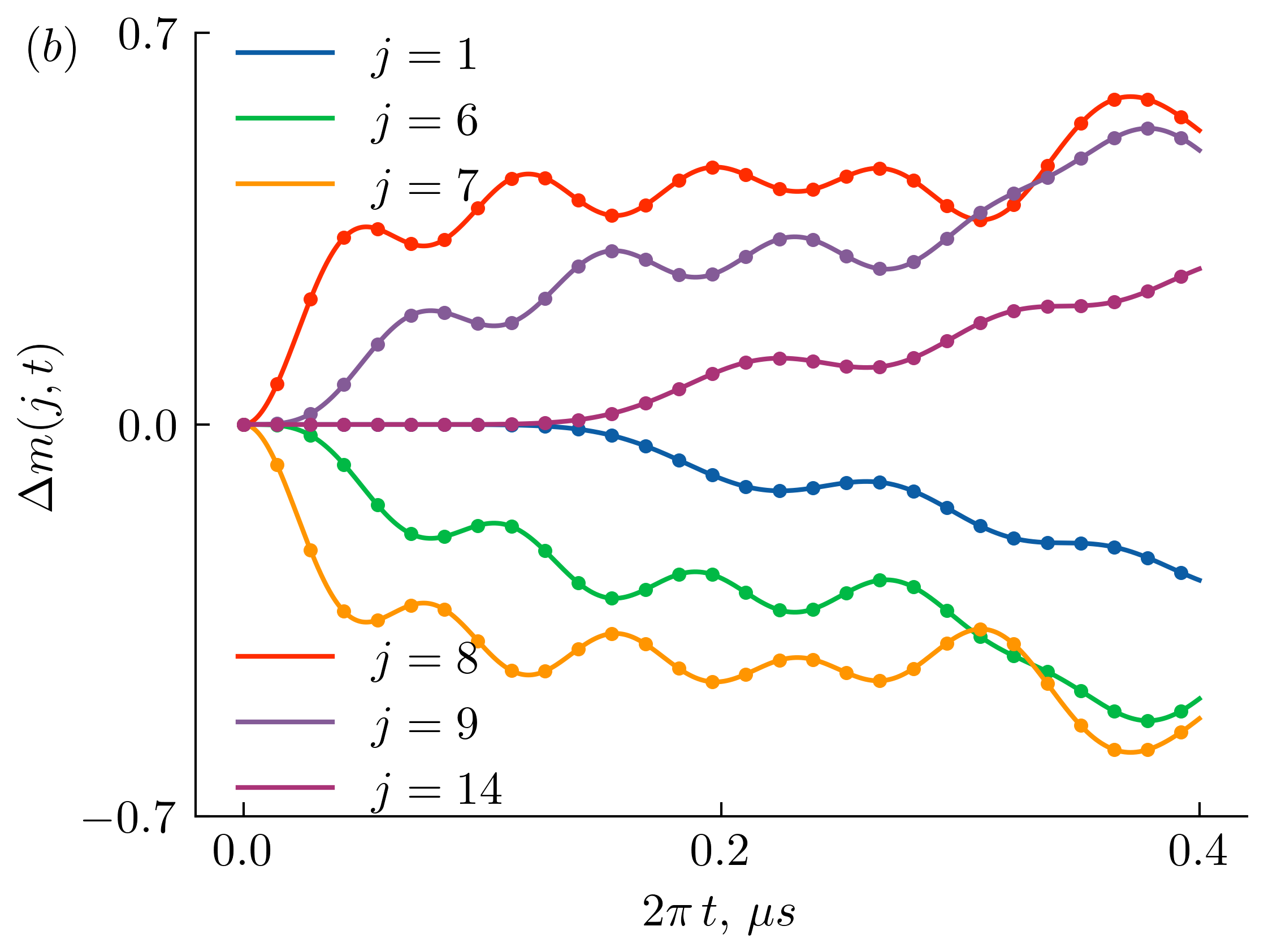}
		\caption{Comparison of the dynamics of the circuit-QED simulator (solid line), Eq. (\ref{eq:H_JJA_simplified}), and the dimerized antiferromagnetic Heisenberg chain with open boundary conditions (dotted line), Eq. (\ref{eq:H1}). The initial state is the domain wall state. The number of sites is $N=14$. The coupling constant is $J=2\pi\hbar\times 40$ MHz. The parameters of the circuit-QED simulator are presented in Table~\ref{tab1}.  (a) Magnetization flow through the center for dimerization strength $\delta = 0,0.1,0.2$, and $0.3$.
(b) Deviations of the spin expectation values from their initial values for sites in the middle ($j = 6, 7, 8, 9$) and at the edges ($j = 1, 14$) of the chain.
The dimerization coefficient is $\delta=0$.  }
\label{fig:magnetization_transport}
	\end{figure*}
    
Fig. \ref{fig:magnetization_transport} illustrates the magnetization transport 
in a spin chain initialized in the domain-wall state. The chain has  $N=14$ sites and  open boundary conditions. As the dimerization strength increases, we observe a slowdown in the growth of the magnetization transferred from the left half to the right half. Furthermore, we see an increasing number of spins involved in the transport. 

Table \ref{tab1} presents a specific realization of the simulator parameters. As detailed in Appendix \ref{appendix:parameter_space}, different sets --- characterized by different 
\(E_{C}/E^{\text{eq}}_{J}\)
  ratios --- can implement the same target spin Hamiltonian. This flexibility applies to all parameter sets listed in Tables \ref{tab2}-\ref{tab4}.
    
\subsection{Spinon dynamics in the  antiferromagnetic Heisenberg chain}
We simulate a one-dimensional Heisenberg chain with constant coupling $J$ and periodic boundary conditions, governed by the Hamiltonian
    \begin{equation}
		\label{eq:H2}
		\hat{H}_{2} =  J\sum_{j=1}^{N}\left(\hat{S}^x_j\hat{S}^x_{j+1}+\hat{S}^y_j\hat{S}^y_{j+1}+\hat{S}^z_j\hat{S}^z_{j+1}\right),
	\end{equation}
    where the number of spins $N$ is even.
Following \cite{vlijm2016spinon},
we focus on the time evolution of the state $\ket{\rm SF_1}_{\text{spin}} = \sqrt{2}\hat{S}^{-}_1\ket{\rm GS}$
resulting from a local spin flip operation $\hat{S}^{-}_1$ on the ground state $\ket{\rm GS}$ of $\hat{H}_2$. In consistency with the Lieb–Mattis theorem \cite{lieb1962ordering}, the total spin of the ground state is zero; hence, it necessarily lies in the zero magnetization sector.
 We examine the dynamics of collective spin excitations (spinons) by computing the equal time nearest-neighbor correlation function, defined as
\begin{eqnarray}\label{eq:correlation}
    \mathcal{C}_{\text{spin}}(j,t) = \langle 
    \hat{S}^z_j\hat{S}^z_{j+1}
    \rangle(t).
\end{eqnarray}

In the bosonic system, the initial state is constructed as $\ket{\rm SF_1}_{\text{boson}} = \sqrt{2}\hat{a}_1\ket{\rm HS}$, where $\ket{\rm HS}$ is the highest energy eigenstate restricted to the subspace of $N/2$ bosonic excitations. This state can be obtained by means of the variational quantum eigensolver (VQE)~\cite{peruzzo2014variational,bharti2022noisy,tilly2022variational,Sapova2022}, a hybrid classical-quantum  optimization algorithm designed for noisy intermediate-scale quantum computers. The VQE algorithm involves a state preparation procedure and measurement of the Hamiltonian expectation value on a quantum computer, followed by optimization techniques on a classical computer.

We now show that the energy of the JJA can be computed experimentally. For the physical states, it reads

\begin{eqnarray}
 \langle\hat{H}_{\rm JJA}\rangle &&= 
\sum_{j=1}^N\left(\hbar\omega_j+\hbar\delta \omega_j-\frac{1}{2}\Delta^{\rm eq}_j\right)\langle\hat{n}_j\rangle\\
     &&+\sum_{\langle j,k\rangle}t_{jk}\langle\hat{a}_{j}^{\dagger}\hat{a}_{k}+\hat{a}_{j}\hat{a}^{\dagger}_{k}\rangle
     -\sum_{\langle j,k\rangle}\Delta_{jk}\langle
    \hat{n}_j\hat{n}_{k}\rangle. \nonumber
\end{eqnarray}
We observe that each term in the above expression can be obtained experimentally: the expectation of the number operators can be computed via the photon
counting technique~\cite{chen2011microwave,peropadre2011approaching, romero2009microwave} while the terms $\langle\hat{a}_{j}^{\dagger}\hat{a}_{k}+\hat{a}_{j}\hat{a}^{\dagger}_{k}\rangle= 2\,\langle \hat{X}_j\hat{X}_{k}\rangle+2\,\langle \hat{P}_j\hat{P}_{k}\rangle$ involving the quadrature operators $\hat{P}_j = \left(\hat{a}_j+\hat{a}_{k}\right)/2i$
can be measured using the homodyne detection~\cite{dicarlo2010preparation,eichler_characterizing_2012}.  

\begin{figure}
		\centering
		\includegraphics[width=0.45\textwidth]{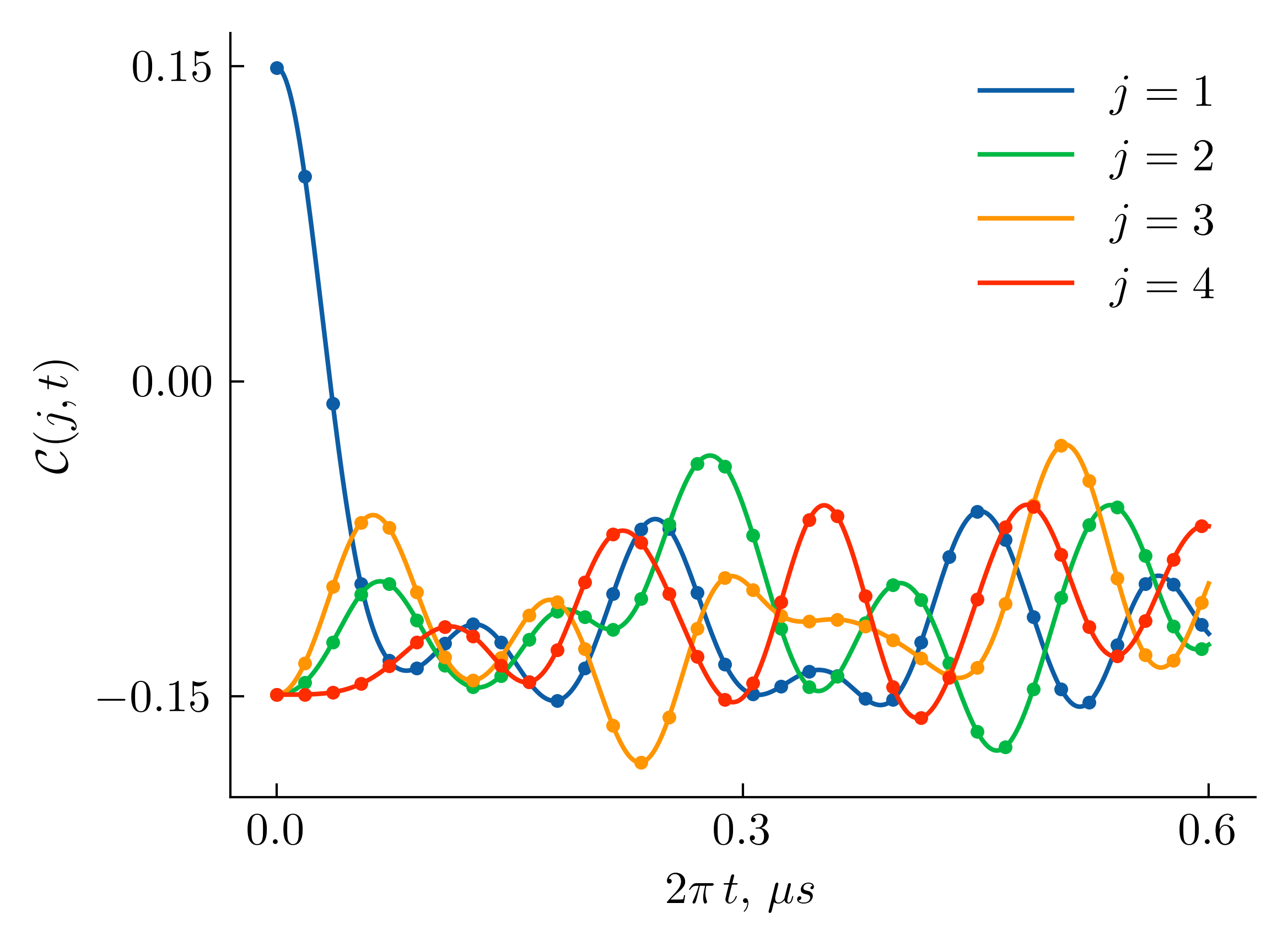}
		\caption{Comparison of the dynamics of the circuit-QED simulator (solid line), Eq. (\ref{eq:H_JJA_simplified}), and the 
         antiferromagnetic Heisenberg chain with periodic boundary conditions (dotted line), Eq. (\ref{eq:H2}). The number of sites is $N=14$. The correlation function is shown for sites $j=1,2,3,4$.
          The coupling constant is $J=2\pi\hbar\times 40$ MHz. The parameters of the circuit-QED simulator are presented in Table~\ref{tab2}. 
         }
    \label{fig:correlation}
	\end{figure} 
    
Fig. \ref{fig:correlation} shows the time evolution of the correlation function for a spin chain with $N=14$ sites and  periodic boundary conditions. It exhibits oscillations after the passage of the spinon.
The parameters of the circuit-QED simulator are presented in Table~\ref{tab2}.

\subsection{Entanglement dynamics in the spatially anisotropic Heisenberg model in two dimensions}
We consider
a spatially anisotropic Heisenberg model on a two-dimensional lattice with open boundary conditions. The model is determined by the Hamiltonian
\begin{equation}
\label{eq:H3}
\hat{H}_3 = J \sum_{\langle j,k\rangle} \bigl(1 + \delta_{jk}\,\eta \bigr)
\left(\hat{S}^x_j\hat{S}^x_k + \hat{S}^y_j\hat{S}^y_k + \hat{S}^z_j\hat{S}^z_k\right).
\end{equation}
The spatial anisotropy parameter is introduced via the bond-direction factor $\delta_{jk}$, with $\delta_{jk} = 1$ for horizontal bonds, $\delta_{jk} = -1$  for vertical bonds, and $\eta$ is the anisotropy parameter.

The system is initialized in the Neel state
$\ket{\rm Neel}_{\text{spin}} = \ket{\uparrow_z\downarrow_z\uparrow_z\downarrow_z\ldots \uparrow_z}$,
in which neighboring spins are aligned in opposite directions. In the bosonic representation, the Neel state corresponds to $\ket{\rm Neel}_{\text{boson}} = \ket{1_10_21_30_4\ldots 0_{N-1}1_N}$. We compute the time evolution of the variance for the staggered magnetization operator
\begin{eqnarray}
    \mathcal{F}(t) = \frac{4}{N}\left(\langle \hat{\mathcal{M}}^2_{\rm stag}\rangle(t)-\langle \hat{\mathcal{M}}_{\rm stag}\rangle(t)^2\right),
        \label{eq:QFI}
\end{eqnarray}
where  $\hat{\mathcal{M}}_{\rm stag, spin}= \sum_{j=1}^N (-1)^j \hat{S}^z_j$ for the spin system and $\hat{\mathcal{M}}_{\rm stag, boson} = \sum_{j=1}^N (-1)^j \left(\hat{n}_j-1/2\right)$ for the bosonic system. This quantity defines the quantum Fisher information density (QFI), which serves as a witness for entanglement (see  Refs.~\cite{morong2021observation,baykusheva2023witnessing}). The condition $\mathcal{F}>1$ guarantees the presence of entanglement in the system. 
\begin{figure} 
        \includegraphics[width=0.9\linewidth]{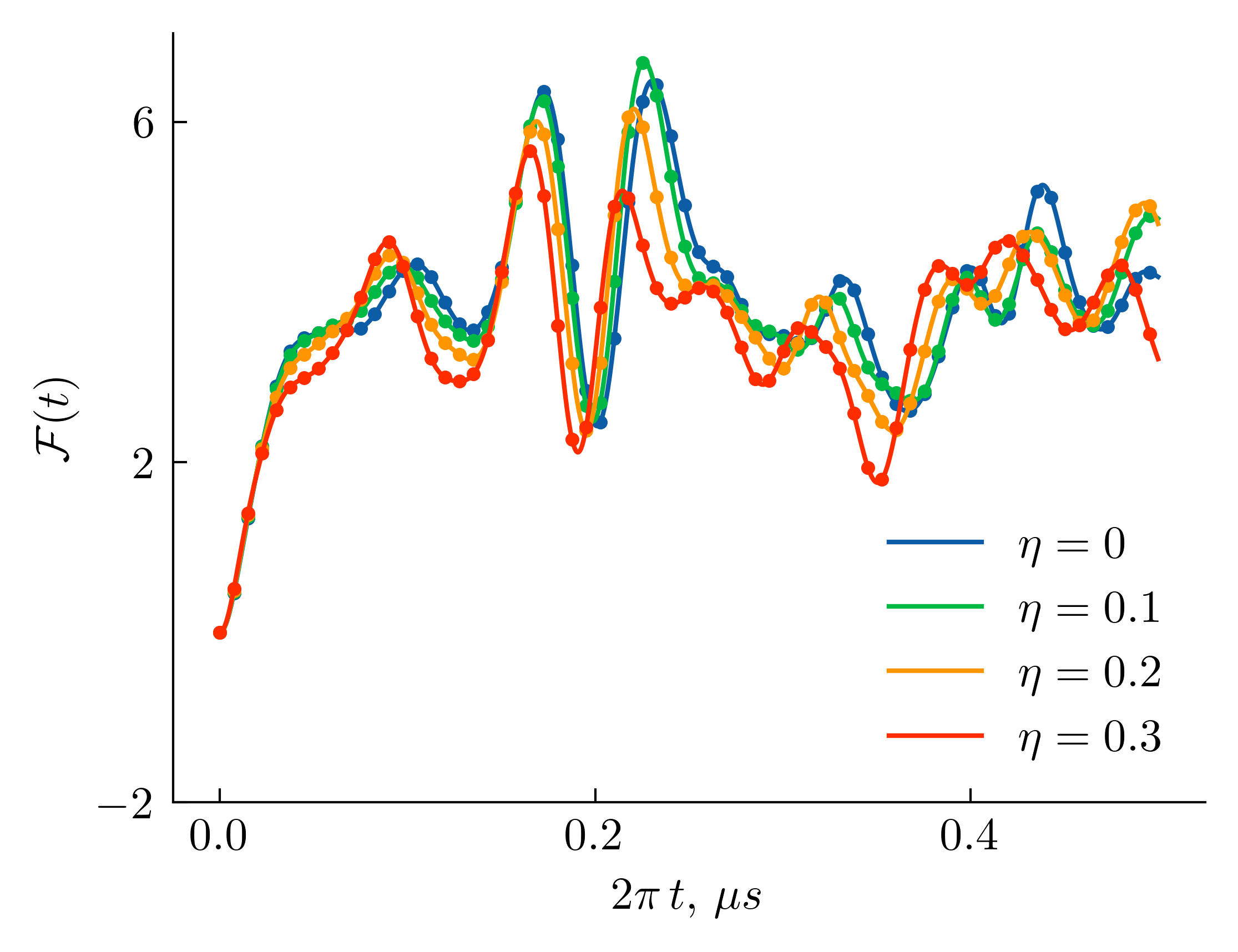}
		\caption{Comparison of the dynamics 
        of the circuit-QED simulator (solid line), Eq. (\ref{eq:H_JJA_simplified}), and 
        the spatially anisotropic Heisenberg model (dotted line), Eq. (\ref{eq:H3}) on a $5\times 3$ lattice with open boundary conditions. The quantum Fisher information density is shown for different anisotropy parameters $\eta = 0, 0.1, 0.2, 0.3$.         
        The coupling constant is $J=2\pi\hbar\times 40$ MHz. The parameters of the circuit-QED simulator are presented in Table~\ref{tab3}.}
		\label{fig:entanglement}
	\end{figure}
    
Fig. \ref{fig:entanglement} presents the dynamics of the QFI
density for a $5\times 3$ spin lattice with open boundary conditions. For all anisotropy parameters considered ($\eta = 0,0.1,0.2$, and $0.3$), the QFI
density grows rapidly and exceeds a value of $1$, signaling entanglement in the system. The parameters of the circuit-QED simulator are presented in Table~\ref{tab3}. 

\subsection{Dynamics of the disordered Heisenberg model in two dimensions}
We use the circuit-QED simulator to reproduce the dynamics of the Heisenberg model on a two-dimensional lattice with open boundary conditions in the presence of disorder. The spin system is described by the Hamiltonian
\begin{equation}
		\label{eq:H4}
		\hat{H}_{4} = - J\sum_{\langle j,k\rangle} \left(\hat{S}^x_j\hat{S}^x_k+\hat{S}^y_j\hat{S}^y_k+\hat{S}^z_j\hat{S}^z_k\right) 
		+\sum_{j=1}^N h_j \hat{S}^z_j.
	\end{equation}
 Here, the local magnetic fields $h_j$ are drawn from the interval $\left(-W,W\right)$, where $W$ is the magnitude of disorder. The initial state is the Neel state. We are interested in the time evolution of the staggered magnetization, also referred to as the imbalance 
\begin{eqnarray}
    \mathcal{I}(t) = \frac{1}{N}\langle\hat{\mathcal{M}}_{\rm stag}\rangle(t).
\end{eqnarray}
The imbalance quantifies the amount of initial-state memory and is used as an indicator of many-body localization (MBL)~\cite{morong2021observation,hur2025stability}.  Specifically, a finite long-time value of the imbalance corresponds to the MBL phase, and its decay to zero corresponds to the thermalized phase.

We note that the existence of the MBL phenomenon in two and higher dimensions remains unclear \cite{krishna2019many,hur2025stability}, largely due to the severe challenges in numerical simulation caused by exponentially increasing complexity with system size.
\begin{figure}
        \includegraphics[width=\linewidth]{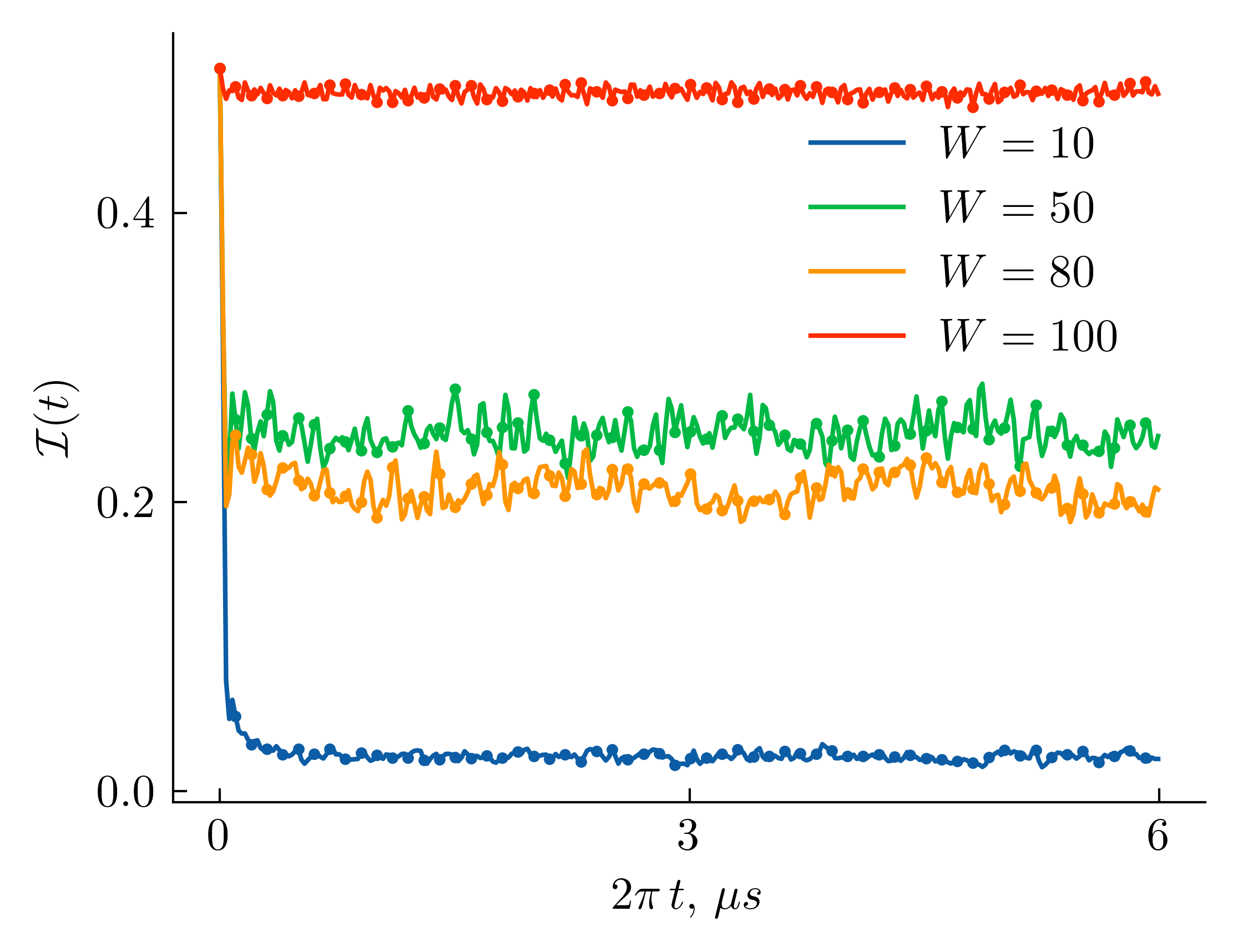}
        \caption{Comparison of the dynamics of
     the circuit-QED simulator (solid line), Eq. (\ref{eq:H_JJA_simplified}), and 
        the disordered Heisenberg model (dotted line), Eq. (\ref{eq:H3}) on a $4\times 4$ lattice with open boundary conditions.  The imbalance is shown for disorder strengths $W=10,50,80$, and $100$.     The coupling constant is $J=2\pi\hbar\times 40$ MHz. The parameters of the circuit-QED simulator are presented in Table~\ref{tab4}.
        }\label{fig:imbalance}
\end{figure}

Fig.\ref{fig:imbalance} depicts the imbalance as a function of time for a $4\times 4$ spin lattice with open boundary conditions. It shows a clear dependence
on the disorder strength $W$, which takes values $W=10,50,80$ and $100$. 
The parameters of the circuit-QED simulator are listed in Table~\ref{tab4}.

The results of our numerical simulations, shown in Figs.~\ref{fig:magnetization_transport}-\ref{fig:imbalance}, confirm that both Hamiltonians yield indistinguishable dynamics (up to machine precision). This agreement conclusively validates the mapping of the spin models to the circuit-QED simulator.   

In the simulations,  we account for the conservation of total magnetization in the spin system. Thus, the dynamics within a fixed magnetization sector is unaffected by the application of a constant external magnetic field.  Moreover, the time evolution of physical quantities is independent of the sign of the coupling constant $J$, as long as the initial state is real.

Additionally, we have verified the equivalence for other observables and initial states, including entangled states and superpositions of states from different magnetization sectors.

\begin{table*}
\caption{\label{tab1} Parameters of the circuit-QED simulator for the numerical simulations in Fig.~\ref{fig:magnetization_transport}, shown for dimerizations $\delta = 0, 0.1, 0.2, 0.3$. For $\delta>0$, parameters differ between strong and weak bonds. We label sites $j=1,N$ as edges and sites $j=2,\ldots,N-1$ as bulk, where $N=14$.}

\begin{center}
\begin{ruledtabular}
\begin{tabular}{c l r r r}
    \multicolumn{5}{c}{\textbf{Sites}} \\
    \hline
    $\delta$ & Site type, $j$ & $\dfrac{\hbar\omega_j}{2\pi\hbar}$ (MHz) &  $\dfrac{E_{C,j}}{2\pi\hbar}$ (MHz) & $\dfrac{E_{J,j}}{2\pi\hbar}$ (MHz) \\[0.5em]
    \hline
    0 & Bulk & 5000 &  200 & 12500 \\
      & Edge & 4958 &  198 & 13930 \\
    \hline
    0.1 & Bulk & 5000 &  200 & 12500 \\
        & Edge & 4954 &  198 & 14075 \\
    \hline
    0.2 & Bulk & 5000 &  200 & 12500 \\
        & Edge & 4950 & 198 & 14218 \\
    \hline
    0.3 & Bulk & 5000 &  200 & 12500 \\
        & Edge & 4946 &  198 & 14363 \\
\end{tabular}
\end{ruledtabular}

\vspace{1em}

\begin{ruledtabular}
\begin{tabular}{c l r r r r r}
    \multicolumn{7}{c}{\textbf{Bonds}} \\
    \hline
    $\delta$ & Bond type, $j-k$ & $\dfrac{t_{jk}}{2\pi\hbar}$ (MHz) & $\dfrac{\Delta_{jk}}{2\pi\hbar}$ (MHz) & $\dfrac{T_{jk}}{2\pi\hbar}$ (MHz) & $\dfrac{E'_{J,jk}}{2\pi\hbar}$ (MHz) & $\dfrac{E_{\mathrm{coup},jk}}{2\pi\hbar}$ (MHz) \\[0.5em]
    \hline
    0 &   & $-$60 & 40 & 20 & 1562 & 30 \\
    \hline
    0.1 & Strong & $-$66 & 44 & 22 & 1719 & 33 \\
        & Weak   & $-$54 & 36 & 18 & 1406 & 27 \\
    \hline
    0.2 & Strong & $-$72 & 48 & 24 & 1875 & 36 \\
        & Weak   & $-$48 & 32 & 16 & 1250 & 24 \\
    \hline
    0.3 & Strong & $-$78 & 52 & 26 & 2031 & 39 \\
        & Weak   & $-$42 & 28 & 14 & 1094 & 21 \\
\end{tabular}
\end{ruledtabular}
\end{center}
\end{table*}

\begin{table*}
\caption{\label{tab2} Parameters of the circuit-QED simulator for the numerical simulations in Fig. \ref{fig:correlation}. The number of sites is $N=14$.}

\begin{center}
\begin{ruledtabular}
\begin{tabular}{c l r r r}
    \multicolumn{5}{c}{\textbf{Sites}} \\
    \hline
    $\delta$ & Site, $j$ & $\dfrac{\hbar\omega_j}{2\pi\hbar}$ (MHz) &  $\dfrac{E_{C,j}}{2\pi\hbar}$ (MHz) & $\dfrac{E_{J,j}}{2\pi\hbar}$ (MHz) \\[0.5em]
    \hline
    0 & 1$-$14 & 5000 &  200 & 12500 \\
\end{tabular}
\end{ruledtabular}

\vspace{1em}

\begin{ruledtabular}
\begin{tabular}{c l r r r r r}
    \multicolumn{7}{c}{\textbf{Bonds}} \\
    \hline
    $\delta$ & Bond, $j-k$ & $\dfrac{t_{jk}}{2\pi\hbar}$ (MHz) & $\dfrac{\Delta_{jk}}{2\pi\hbar}$ (MHz) & $\dfrac{T_{jk}}{2\pi\hbar}$ (MHz) & $\dfrac{E'_{J,jk}}{2\pi\hbar}$ (MHz) & $\dfrac{E_{\mathrm{coup},jk}}{2\pi\hbar}$ (MHz) \\[0.5em]
    \hline
    0 &   & $-$60 & 40 & 20 & 1562 & 30 \\
\end{tabular}
\end{ruledtabular}
\end{center}
\end{table*}

\begin{table*}
\caption{\label{tab3} Parameters of the circuit-QED simulator for the numerical simulations in Fig. \ref{fig:entanglement}, shown for anisotropy parameters $\eta = 0, 0.1, 0.2, 0.3$. The rectangular lattice has a $5\times3$ geometry, with sites enumerated row-wise from the top-left to bottom-right corner. The lattice sites are partitioned into bulk ($j =7-9$), edges ($j=2- 4, 12-14$ for top/bottom and $j=6,10$ for left/right) and corners ($j=1,5,11,15$). For $\eta>0$, parameters differ between horizontal and vertical bonds.}

\begin{center}
\begin{ruledtabular}
\begin{tabular}{c l r r r}
    \multicolumn{5}{c}{\textbf{Sites}} \\
    \hline
    $\eta$ & Site type, $j$ & $\dfrac{\hbar\omega_j}{2\pi\hbar}$ (MHz) & $\dfrac{E_{C,j}}{2\pi\hbar}$ (MHz) & $\dfrac{E_{J,j}}{2\pi\hbar}$ (MHz) \\[0.5em]
    \hline
    0 & Bulk & 5082 & 202 & 9658 \\
      & Edges & 5040 & 200 & 11114 \\
      & Corners & 4999 & 198 & 12573 \\
    \hline
    0.1 & Bulk & 5082 & 202 & 9658 \\
        & Edges, top/bottom & 5044 & 200 & 10968 \\
        & Edges, left/right & 5036 & 200 & 11260 \\
        & Corners & 4998 & 198 & 12570 \\
    \hline
    0.2 & Bulk & 5082 & 202 & 9658 \\
        & Edges, top/bottom & 5048 & 200 & 10822 \\
        & Edges, left/right & 5032 & 200 & 11406 \\
        & Corners & 4998 & 198 & 12570 \\
    \hline
    0.3 & Bulk & 5082 & 202 & 9652 \\
        & Edges, top/bottom & 5053 & 201 & 10668 \\
        & Edges, left/right & 5028 & 200 & 11539 \\
        & Corners & 4999 & 199 & 12555 \\
\end{tabular}
\end{ruledtabular}

\vspace{1em}

\begin{ruledtabular}
\begin{tabular}{c l r r r r r}
    \multicolumn{7}{c}{\textbf{Bonds}} \\
    \hline
    $\eta$ & Bond type, $j-k$ & $\dfrac{t_{jk}}{2\pi\hbar}$ (MHz) & $\dfrac{\Delta_{jk}}{2\pi\hbar}$ (MHz) & $\dfrac{T_{jk}}{2\pi\hbar}$ (MHz) & $\dfrac{E'_{J,jk}}{2\pi\hbar}$ (MHz) & $\dfrac{E_{\mathrm{coup},jk}}{2\pi\hbar}$ (MHz) \\[0.5em]
    \hline
    0 &  & $-$60 & 40 & 20 & 1588 & 30 \\
    \hline
    0.1 & Horizontal & $-$66 & 44 & 22 & 1747 & 34 \\
        & Vertical & $-$54 & 36 & 18 & 1429 & 27 \\
    \hline
    0.2 & Horizontal & $-$72 & 48 & 24 & 1906 & 37 \\
        & Vertical & $-$48 & 32 & 16 & 1270 & 24 \\
    \hline
    0.3 & Horizontal & $-$78 & 52 & 26 & 1950 & 39 \\
        & Vertical & $-$42 & 28 & 14 & 1050 & 21 \\
\end{tabular}
\end{ruledtabular}
\end{center}
\end{table*}

\begin{table*}
\caption{\label{tab4} Parameters of the circuit-QED simulator for the numerical simulations in Fig. \ref{fig:imbalance}. For each disorder strength $W = 10, 50, 80, 100$ of the spin lattice, the table provides the range of the circuit-QED simulator parameters.}

\begin{center}
\begin{ruledtabular}
\begin{tabular}{c r r r}
    \multicolumn{4}{c}{\textbf{Site Parameters }} \\
    \hline
    $W$ & $\dfrac{\hbar\omega_j}{2\pi\hbar}$ (MHz) & $\dfrac{E_{C,j}}{2\pi\hbar}$ (MHz) & $\dfrac{E^{}_{J,j}}{2\pi\hbar}$ (MHz) \\[0.5em]
    \hline
    10 & $4994$--$5098$ & $204$--$208$ & $9546$--$12354$ \\
    \hline
    50 & $4952$--$5140$ & $202$--$210$ & $9418$--$12482$ \\
    \hline
    80 & $4931$--$5140$ & $201$--$210$ & $9386$--$12578$ \\
    \hline
    100 & $4827$--$5244$ & $197$--$214$ & $9035$--$12673$ \\
\end{tabular}
\end{ruledtabular}

\vspace{1em}

\begin{ruledtabular}
\begin{tabular}{c r r r r r}
    \multicolumn{6}{c}{\textbf{Bond Parameters}} \\
    \hline
    $W$ & $\dfrac{t_{jk}}{2\pi\hbar}$ (MHz) & $\dfrac{\Delta_{jk}}{2\pi\hbar}$ (MHz) & $\dfrac{T_{jk}}{2\pi\hbar}$ (MHz) & $\dfrac{E'_{J,jk}}{2\pi\hbar}$ (MHz) & $\dfrac{E_{\mathrm{coup},jk}}{2\pi\hbar}$ (MHz) \\[0.5em]
    \hline
    \multicolumn{6}{c}{} \\
    10--100 & $-$60 & 40 & 20 & 1500 & 30 \\
\end{tabular}
\end{ruledtabular}
\end{center}
\end{table*}

\section{discussion\label{sec:discussion}}

    The classical simulation of real-time dynamics for one- and two-dimensional Heisenberg models is computationally intractable at intermediate scales ($N \sim 50\text{--}100$ spins). Exact diagonalization is limited to system sizes of $N \sim 20\text{--}30$ \cite{sandvik2010computational,PhysRevB.74.020403} spins by the exponential growth of the Hilbert space. Tensor-network methods (such as matrix product states and projected entangled pair states) are constrained by the growth of entanglement entropy \cite{Calabrese_2005}, which restricts simulations to relatively short time scales \cite{PhysRevLett.93.040502,ORUS2014117}. Furthermore, quantum monte carlo techniques, while powerful for equilibrium properties \cite{RevModPhys.73.33,sandvik2010computational}, suffer from the sign problem for real-time evolution \cite{PhysRevLett.94.170201,PhysRevLett.115.266802}. 
A promising alternative is the neural quantum states approach \cite{Lange_2024}, which has demonstrated the capability to simulate systems of up to $N \sim 100$ spins \cite{carleo2017solving}. However, achieving high accuracy faces a significant challenge in the optimizing of its high-dimensional neural network parameters \cite{Lange_2024}.

To circumvent the limitations of classical simulations, in this study, we have proposed a circuit quantum electrodynamics (circuit-QED) simulator for the Heisenberg spin model. Using the Dyson-Maleev transformation,we have mapped the spin system to the extended Bose-Hubbard (EBH) model — a bosonic Hamiltonian with parameters that are naturally engineered within a Josephson-junction array. This makes the proposed simulator realistically implementable with current superconducting circuit technology~\cite{Oliver2019,RamosJJA}.

A key feature of the EBH system is that its dynamics is restricted to the subspace of two physical states per site. This constraint guarantees the equivalence between the EBH model and the original Heisenberg spin system.
To verify this equivalence, we have performed numerical simulations. The results demonstrate excellent agreement between the evolution of spin$-1/2$ observables and the corresponding microwave-photon dynamics in one- and two-dimensional lattices. Consequently, the proposed simulator provides a scalable pathway for exploring the dynamics of quantum spin systems beyond the reach of current classical computation methods.

Our approach, based on the Dyson-Maleev transformation, corresponds to the Holstein-Primakoff encoding for spin$-1/2$ systems, which ensures its independence from the specific bosonic representation used.  

The experimental realization of the proposed circuit-QED simulator requires the adjustment of parameters to within a few MHz. Although challenging, this is feasible for modern superconducting circuit technology. 
Inherent variations in the fabrication of Josephson junctions and capacitors typically result in a parameter spread of a few percent. The required precision can be achieved through local flux addressability, which enables individual parameter tuning.
Furthermore, the mapping to the circuit-QED system provides significant calibration flexibility, 
as the same spin Hamiltonian can be realized by multiple parameter configurations.
However, inaccuracies of a few MHz will cause the simulated dynamics to deviate from those of the target Heisenberg spin model at long times.

A promising direction for future research is to explore high-spin systems (see~\cite{Martinis2014,Kiktenko2015,Kiktenko20152,Ustinov2015,Zinner2019,Roy2023,Blok2025,Fedorov2025}), which may be utilized to simulate exotic models relevant to various physical systems~\cite{Ringbauer2024,Carretta2024}. 
In the context of the dimerized chain model studied here, an immediate next step would be to move beyond the adiabatic approximation by treating the lattice degrees of freedom dynamically and thus recover the full spin-Peierls Hamiltonian.
Furthermore, our framework could be extended to engineer couplings beyond nearest neighbors and to study dynamics of open quantum systems  via programmable cross-Kerr networks.

	
	\begin{acknowledgments} 
		We thank Juan Jos\'e Garc\'ia Ripoll for the in-depth  discussion.
A.K.F. thanks the Priority 2030 program at the NUST ``MISIS'' under the project K1-2022-027.
	J.H. acknowledges the support of the following multiple funding sources:
[1] Basic Science Research Program through the National Research Foundation of Korea (NRF), funded by the Ministry of Science and ICT (RS-2023-NR068116, RS-2023-NR119931, RS-2025-03532992). [2] Institute for Information \& Communications Technology Promotion (IITP) grant funded by the Korea government (MSIP) (No. 2019-0-00003), which focuses on the research and development of core technologies for programming, running, implementing, and validating fault-tolerant quantum computing systems. [3] Yonsei University Research Fund under project number 2025-22-0140. T.R. acknowledges funding from the Generaci\'on de Conocimientos project PID2023-146531NA-I00 and the Ram\'on y Cajal program RYC2021-032473-I, financed by MCIN/AEI/10.13039/501100011033 and the European Union NextGenerationEU/PRTR.

	\end{acknowledgments}
    
\section*{Data Availability}
The data that generate Figs. \ref{fig:magnetization_transport} -- \ref{fig:imbalance} are openly available \cite{spin_dynamics_data_2025}. 

\section*{Appendix}
    \appendix
 
	\section{Hamiltonian of the Josephson junction array\label{appendix:H_JJA derivation}}
	
	In this Appendix, we present the detailed derivation of the Hamiltonian of the superconducting circuit given by Eq.~(\ref{eq:JJA_Ham}). The system is described by the Lagrangian of the following form \cite{RamosJJA}:
	\begin{eqnarray}
		\label{eq:lagrangian}
		{\cal L}_{\rm JJA}&=& {}\sum_{j=1}^N \frac{C_j(\dot{\phi}_j)^2}{2}
		+\sum_{\langle j,k\rangle}\frac{C'_{jk}}{2}(\dot{\phi}_{j}-
		\dot{\phi}_{k})^2\\
		&+&\sum_{j=1}^N E_{J,j}\cos\left(\frac{\phi_j}{\Phi_0}\right)+\sum_{\langle j,k\rangle}E'_{J,jk}\cos\left(\frac{\phi_{j}-\phi_{k}}{\Phi_0}\right),\nonumber
	\end{eqnarray}
	where ${\phi}_j$,  $j=1,\ldots,N$, are the flux variables. Here, double sums are performed over connected sites of a lattice.
	
	In order to quantize the system, we define the conjugate charge variables
    \begin{eqnarray}
		q_j &=& \partial{\cal L_{\rm JJA}}/\partial\dot{\phi}_j =
        C_j\,\dot{\phi}_{j}+\sum_{k\in \text{NN}(j)}C'_{jk}\left(\dot{\phi}_j-\dot{\phi}_k\right)\nonumber\\&=&
		C^{\text{eq}}_j\,\dot{\phi}_{j}-\sum_{k\in \text{NN}(j)}C'_{jk}\dot{\phi}_k,
	\end{eqnarray}
    where the index $j$ runs from $1$ to $N$ and the sum is taken over the nearest-neighbors of the $j$-th lattice site. We introduce the equivalent capacitances as 
    \begin{equation}
        C^{\text{eq}}_j = C_j + \sum_{k\in \text{NN}(j)}C'_{jk}.
    \end{equation}
	Under the assumption of low coupling capacitances, $C'_{jk} \ll C_j$, the inverse transform is given by
	\begin{eqnarray}
		\dot{\phi}_j=A_j \,q_j + \sum_{k\in \text{NN}(j)}A'_{jk} \,q_{k},
	\end{eqnarray}
	where the coefficients are defined by
	\begin{eqnarray}
		A_j &=& \frac{1}{C^{\text{eq}}_j},\label{app:Aj}\\
		A'_{jk}&=&\frac{C'_{jk}}{C^{\text{eq}}_j\,C^{\text{eq}}_k}\label{app:Ajk}.
	\end{eqnarray}
	We determine the Hamiltonian of the circuit by the Legendre transformation. To do this, we express the kinetic energy term through the charge variables as 
	\begin{eqnarray}
		{\cal L}_{\rm kin} &&= {}\sum_{j=1}^N \frac{C_j(\dot{\phi}_j)^2}{2}
		+\sum_{\langle j,k\rangle}\frac{C'_{jk}}{2}(\dot{\phi}_{j}-
		\dot{\phi}_{k})^2\\
		&&={}\sum_{j=1}^N \frac{C_j}{2}\left( A_j \,q_j + \sum_{k\in \text{NN}(j)}A'_{jk} \,q_{k}\right)^2\nonumber\\
		&&+\sum_{\langle j,k\rangle}\frac{C'_{jk}}{2}\left(A_j \,q_j-A_k\,q_k\right)^2 \nonumber \\
        &&= {}\sum_{j=1}^N \frac{C_j}{2} A^2_j \,q^2_j+\sum_{j=1}^N\sum_{k\in \text{NN}(j)}C_j\,A_j\,A'_{jk}\,q_j\,q_k\nonumber\\
        &&+\sum_{\langle j,k\rangle}\frac{C'_{jk}}{2}\left(A^2_j \,q^2_j+A^2_k\,q^2_k\right)-\sum_{\langle j,k\rangle}C'_{jk}\,A_j\,A_k\, q_j\,q_k,
	\nonumber
    \end{eqnarray}
    where we have neglected small terms. Regrouping the terms,
	we obtain
	\begin{eqnarray}
	    {\cal L}_{\rm kin}&&={}\sum_{j=1}^N \frac{C_j}{2} A^2_j \,q^2_j+\sum_{\langle j,k\rangle}q_j\,q_k\,A'_{jk}\left(C_j\,A_j+C_k\,A_k\right)\nonumber\\
        &&+\sum_{j=1}^N\sum_{k\in \text{NN}(j)}\frac{C'_{jk}}{2}A^2_j \,q^2_j-\sum_{\langle j,k\rangle}C'_{jk}\,A_j\,A_k\, q_j\,q_k\nonumber\\
        &&=\sum_{\langle j,k\rangle}q_j\,q_k\left[A'_{jk}\left(C_j\,A_j+C_k\,A_k\right)-C'_{jk}\,A_j\,A_k\right]\nonumber\\
        &&+\sum_{j=1}^N\frac{C^{\text{eq}}_j}{2} A^2_j \,q^2_j.
	\end{eqnarray}
    Using Eqs. (\ref{app:Aj}) and (\ref{app:Ajk}), we arrive at the final expression for the kinetic energy 
       \begin{eqnarray}
		{\cal L}_{\rm kin}&&=\sum_{j=1}^N\frac{q^2_j}{2\,C^{\text{eq}}_j}+\sum_{\langle j,k\rangle}q_j\,q_k\frac{C'_{jk}}{C^{\text{eq}}_j\,C^{\text{eq}}_k}\left(\frac{C_j}{C^{\text{eq}}_j}+\frac{C_k}{C^{\text{eq}}_k}-1\right)\nonumber\\
        &&=\sum_{j=1}^N\frac{q^2_j}{2\,C^{\text{eq}}_j}+\sum_{\langle j,k\rangle}q_j\,q_k\frac{C'_{jk}}{C^{\text{eq}}_j\,C^{\text{eq}}_k}\nonumber\\
        &&=\sum_{j=1}^N E_{C,j}\,\frac{q_j^2}{e^2}+\sum_{\langle j,k\rangle}E_{\text{coup},jk}\,\frac{q_j\,q_{k}}{e^2},
    \end{eqnarray}
	where the capacitive energies $E_{C,j}$ and capacitive coupling energies $E_{{\text{coup}},j}$ are given by
	\begin{eqnarray}
		E_{C,j} &=&\frac{e^2}{2\,C^{\text{eq}}_j},\label{eq_appendix:EC}\\
		E_{{\rm coup},jk} &=&\frac{e^2\,C'_{jk}}{C^{\text{eq}}_j\,C^{\text{eq}}_k}\label{eq_appendix:Ecoup}.
	\end{eqnarray}
	The classical Hamiltonian of the system has the form 
	\begin{eqnarray}
		{H}_{\rm JJA}&&=\sum_{j=1}^N E_{C,j}\,\frac{{q}_j^2}{e^2}+\sum_{\langle j,k\rangle}E_{\text{coup},jk}\,\frac{q_j\,q_{k}}{e^2} \\&&
		-\sum_{j=1}^N E_{J,j}\cos\left(\frac{{\phi}_j}{\Phi_0}\right)
		-\sum_{\langle j,k\rangle}E'_{J,jk}\cos\left(\frac{{\phi}_{j}-{\phi}_{k}}{\Phi_0}\right).\nonumber
	\end{eqnarray}
	We now quantize the classical Hamiltonian by replacing the flux and charge variables with the operators that satisfy the canonical commutation relation $[q_j, \phi_k]=i\hbar\delta_{j, k}$, where $j,k=1,\ldots,N.$ The quantum Hamiltonian of the circuit becomes
	\begin{eqnarray}
		\label{eq:lagrangian}
		\hat{H}_{\rm JJA}&&=\sum_{j=1}^N E_{C,j}\,\frac{\hat{q}_j^2}{e^2}+\sum_{\langle j,k\rangle}E_{\text{coup},jk}\,\frac{\hat{q}_j\,\hat{q}_{k}}{e^2} \\&&
		-\sum_{j=1}^N E_{J,j}\cos\left(\frac{\hat{\phi}_j}{\Phi_0}\right)
		-\sum_{\langle j,k\rangle}E'_{J,jk}\cos\left(\frac{\hat{\phi}_{j}-\hat{\phi}_{k}}{\Phi_0}\right).\nonumber
	\end{eqnarray}

    Let us now consider the potential term of the Hamiltonian. We assume that the capacitive energies, \( E_{C,j} \), are much smaller than the Josephson energies, \( E_{J,jk} \). In this regime, the fluctuations satisfy \( \langle \hat{\phi}_j^2 \rangle \ll \Phi_0^2 \) and \( \langle (\hat{\phi}_j - \hat{\phi}_k)^2 \rangle \ll \Phi_0^2 \). Therefore, the cosine functions can be expanded, yielding the following approximation for the potential term up to fourth order in the flux:
	\begin{eqnarray}
		\hat{U}_{\rm pot}&=&-
		\sum_{j=1}^N E_{J,j}\cos\left(\frac{\hat{\phi}_j}{\Phi_0}\right)
		-\sum_{\langle j,k\rangle}E'_{J,jk}\cos\left(\frac{\hat{\phi}_{j}-\hat{\phi}_{k}}{\Phi_0}\right)\nonumber\\&&
		=
		\sum_{j=1}^N E_{J,j}\left(\frac{\hat{\phi}^2_j}{2\,\Phi^2_0}-\frac{\hat{\phi}^4_j}{24\,\Phi^4_0}\right)\\&&+
		\sum_{\langle j,k\rangle} E'_{J,jk}\left(\frac{(\hat{\phi}_j-\hat{\phi}_{k})^2}{2\,\Phi^2_0}-\frac{(\hat{\phi}_j-\hat{\phi}_{k})^4}{24\,\Phi^4_0}\right).\nonumber
	\end{eqnarray}
    Terms that are quadratic in the flux read
    \begin{eqnarray}
       &&\sum_{j=1}^N E_{J,j}\frac{\hat{\phi}^2_j}{2\,\Phi^2_0}+
        \sum_{\langle j,k\rangle}E'_{J,jk}\frac{\hat{\phi}^2_j+\hat{\phi}^2_k}{2\,\Phi^2_0} \\&&= \sum_{j=1}^N \frac{\hat{\phi}^2_j}{2\,\Phi^2_0}\left( E_{J,j}+\sum_{k\in \text{NN}(j)} E'_{J,jk}\right)=\sum_{j=1}^N \frac{\hat{\phi}^2_j}{2\,\Phi^2_0}E^{\text{eq}}_{J,j},\nonumber
    \end{eqnarray}
    where we define the equivalent Josephson energies as
    \begin{eqnarray}
E^{\text{eq}}_{J,j}=E_{J,j}+\sum_{k\in \text{NN}(j)} E'_{J,jk}.\label{eq_appendix:EL}
    \end{eqnarray}
Similarly, we regroup other terms  to obtain the following expression
     \begin{eqnarray}
       &&-\sum_{j=1}^N E_{J,j}\frac{\hat{\phi}^4_j}{24\,\Phi^4_0}-
        \sum_{\langle j,k\rangle}E'_{J,jk}\frac{\hat{\phi}^4_j+\hat{\phi}^4_k}{24\,\Phi^4_0} =-\sum_{j=1}^N \frac{\hat{\phi}^4_j}{24\,\Phi^4_0}E^{\text{eq}}_{J,j}\nonumber
    \end{eqnarray}
    The potential energy term reads
	\begin{eqnarray}
		\hat{U}_{\rm pot}&=&\sum_{j=1}^N \frac{\hat{\phi}^2_j}{2\,\Phi^2_0}E^{\text{eq}}_{J,j}-\sum_{j=1}^N \frac{\hat{\phi}^4_j}{24\,\Phi^4_0}E^{\text{eq}}_{J,j}\nonumber\\
		&-&
		\sum_{\langle j,k\rangle} E'_{J,jk}\,\frac{\hat{\phi}_j\,\hat{\phi}_{k}}{\Phi^2_0} +\sum_{\langle j,k\rangle} E'_{J,jk}\frac{\hat{\phi}_j\hat{\phi}_{k}}{6\,\Phi^4_0}\left(\hat{\phi}^2_j+\hat{\phi}^2_{k}\right)
		\nonumber\\&-&\sum_{\langle j,k\rangle} E'_{J,jk}\frac{\hat{\phi}^2_j\hat{\phi}^2_{k}}{4\,\Phi_0^4},   
	\end{eqnarray}
	The Hamiltonian of the circuit takes the form 
	\begin{eqnarray}
		\hat{H}_{\rm JJA}&=&\sum_{j=1}^N \left(E_{C,j}\,\frac{\hat{q}_j^2}{e^2}
		+E^{\text{eq}}_{J,j}\frac{\hat{\phi}_j^2}{2\,\Phi^2_0}\right) -\sum_{j=1}^NE^{\text{eq}}_{J,j}\frac{\hat{\phi}_j^4}{24\,\Phi^4_0}\nonumber\\
		&+&\sum_{\langle j,k \rangle}\left(E_{\text{coup},jk}\,\frac{\hat{q}_j\,\hat{q}_{k}}{e^2}-E'_{J,jk}\,\frac{\hat{\phi}_j\,\hat{\phi}_{k}}{\Phi^2_0}\right)\\
		&-&\sum_{\langle j,k \rangle} E'_{J,jk}\,\frac{\hat{\phi}_{j}^2\,\hat{\phi}_{k}^2}{4\,\Phi^4_0}
		+\sum_{\langle j,k \rangle} E'_{J,jk}\,\frac{\hat{\phi}_{j}\,\hat{\phi}_{k}}{6\,\Phi^4_0}\left(\hat{\phi}_{j}^2+\hat{\phi}_{k}^2\right).\nonumber
	\end{eqnarray}
	We express the Hamiltonian 
	in terms of the creation $\hat{a}^{\dagger}_j$ and annihilation $\hat{a}_j$ operators defined by equations
	\begin{eqnarray}
		\frac{\hat{\phi}_j}{\Phi_0} &=& \left(\frac{2E_{C,j}}{E^{\rm eq}_{J,j}}\right)^{1/4}\left(\hat{a}_j+\hat{a}^{\dagger}_j\right), \\ 
		\frac{\hat{q}_j}{e} &=& i\left(\frac{E^{\rm eq}_{J,j}}{2E_{C,j}}\right)^{1/4}
		\left(\hat{a}^{\dagger}_j-\hat{a}_j\right). 
	\end{eqnarray}
	The first part of the Hamiltonian describes the system of $N$ linear harmonic oscillators 
	\begin{eqnarray}
		&&\sum_{j=1}^N \left(E_{C,j}\,\frac{\hat{q}_j^2}{e^2}
		+E^{\rm eq}_{J,j}\frac{\hat{\phi}_j^2}{2\,\Phi^2_0}\right)\\&&=
		-\sum_{j=1}^N  E_{C,j}\,\left(\frac{E^{\rm eq}_{J,j}}{2E_{C,j}}\right)^{1/2}
		\left(\hat{a}^{\dagger}_j-\hat{a}_j\right)^2
		\nonumber\\&&+\sum_{j=1}^{N}\frac{E^{\rm eq}_{J,j}}{2}\left(\frac{2E_{C,j}}{E^{\rm eq}_{J,j}}\right)^{1/2}\left(\hat{a}_j+\hat{a}^{\dagger}_j\right)^2\nonumber\\&&=\sum_{j=1}^{N}\sqrt{2E_{C,j}E^{\rm eq}_{J,j}}\left(\hat{a}_{j}^{\dagger}\hat{a}_{j}+\hat{a}_{j}\hat{a}^{\dagger}_{j}\right)=\sum_{j=1}^{N}\hbar\,\omega_j\hat{a}_{j}^{\dagger}\hat{a}_{j},\nonumber
	\end{eqnarray}
where we remove all constant terms and use the rotating-wave approximation (RWA) in which the terms that do not conserve the number of excitations are omitted. The bare oscillator frequencies are given by
	\begin{eqnarray}
		\omega_j = \sqrt{8E_{C,j}E^{\rm eq}_{J,j}}/\hbar.\label{eq:frequency}
	\end{eqnarray}
	The second part of the Hamiltonian contributes to the frequency correction
	\begin{eqnarray}
		&&-\sum_{j=1}^NE^{\rm eq}_{J,j}\frac{\hat{\phi}_j^4}{24\,\Phi^4_0} = 
		-\sum_{j=1}^N \frac{E^{\rm eq}_{J,j}}{24}\frac{2E_{C,j}}{E^{\rm eq}_{J,j}}\left(\hat{a}_j+\hat{a}^{\dagger}_j\right)^4 \nonumber\\&&=
		-\sum_{j=1}^N E_{C,j}\left(\hat{a}^{\dagger}_j\hat{a}_j +\frac{1}{2}(\hat{a}^{\dagger}_j)^2\hat{a}^2_j\right)\nonumber\\&&=
		\sum_{j=1}^N \hbar\,\delta\omega_j\left(\hat{a}^{\dagger}_j\hat{a}_j +\frac{1}{2}(\hat{a}^{\dagger}_j)^2\hat{a}^2_j\right),
	\end{eqnarray}
	where $\delta \omega_j$ refer to the anharmonicity parameters
	\begin{eqnarray}
		\delta\omega_j = -E_{C,j}/\hbar.
	\end{eqnarray}
	The third part of the Hamiltonian represents the linear coupling between the oscillators
	\begin{eqnarray}
		&&\sum_{\langle j,k\rangle}\left(E_{\text{coup},jk}\,\frac{\hat{q}_j\,\hat{q}_{k}}{e^2}-E'_{J,jk}\,\frac{\hat{\phi}_j\,\hat{\phi}_{k}}{\Phi^2_0}\right) \nonumber\\&&=
		\sum_{\langle j,k\rangle}-E_{\text{coup},jk}\left(\frac{E^{\rm eq}_{J,j}}{2E_{C,j}}\right)^{1/4}
		\left(\frac{E^{\rm eq}_{J,k}}{2E_{C,k}}\right)^{1/4}\nonumber\\
		&&\times\left(\hat{a}^{\dagger}_j-\hat{a}_j\right)
		\left(\hat{a}^{\dagger}_{k}-\hat{a}_{k}\right)
		-\sum_{\langle j,k\rangle}
		E'_{J,jk} \left(\frac{2E_{C,j}}{E^{\rm eq}_{J,j}}\right)^{1/4}
		\nonumber\\
		&&\times\left(\frac{2E_{C,k}}{E^{\rm eq}_{J,k}}\right)^{1/4}
		\left(\hat{a}^{\dagger}_j+\hat{a}_j\right)
		\left(\hat{a}^{\dagger}_{k}+\hat{a}_{k}\right).
	\end{eqnarray}
	Performing the RWA, we have
	\begin{eqnarray}
		&&\sum_{\langle j,k\rangle}E_{\text{coup},jk}\left(\frac{E^{\rm eq}_{J,j}}{2E_{C,j}}\right)^{1/4}
		\left(\frac{E^{\rm eq}_{J,k}}{2E_{C,k}}\right)^{1/4}
		\left(\hat{a}^{\dagger}_j\hat{a}_{k}+\hat{a}_j\hat{a}^{\dagger}_{k}\right)\nonumber\\
		&&-\sum_{\langle j,k\rangle}
		E'_{J,jk} \left(\frac{2E_{C,j}}{E^{\rm eq}_{J,j}}\right)^{1/4}
		\left(\frac{2E_{C,k}}{E^{\rm eq}_{J,k}}\right)^{1/4}
		\left(\hat{a}^{\dagger}_j\hat{a}_{k}+\hat{a}_j\hat{a}^{\dagger}_{k}\right)\nonumber\\&&=
		\sum_{\langle j,k\rangle}t_{jk}\left(\hat{a}_{j}^{\dagger}\hat{a}_{k}+\hat{a}_{j}\hat{a}^{\dagger}_{k}\right),
	\end{eqnarray}
	where we define the linear hopping parameters as
	\begin{eqnarray}
		t_{jk} &=& E_{\text{coup},jk}\left(\frac{E^{\rm eq}_{J,j}}{4E_{C,j}}\frac{E^{\rm eq}_{J,k}}{E_{C,k}}\right)^{1/4}\nonumber\\&&-E'_{J,jk}\left(\frac{4E_{C,j}}{E^{\rm eq}_{J,j}}\frac{E_{C,k}}{E^{\rm eq}_{J,k}}\right)^{1/4}.
	\end{eqnarray}
	The next part of the Hamiltonian is associated with the cross-Kerr couplings
	\begin{eqnarray}
		&&-\sum_{\langle j,k\rangle} E'_{J,jk}\,\frac{\hat{\phi}_{j}^2\,\hat{\phi}_{k}^2}{4\,\Phi^4_0}=
		-\sum_{\langle j,k\rangle}\frac{E'_{J,jk}}{4}
		\left(\frac{2E_{C,j}}{E^{\rm eq}_{J,j}}\frac{2E_{C,k}}{E^{\rm eq}_{J,k}}\right)^{1/2}\nonumber\\&&\times
		\left(\hat{a}_j+\hat{a}^{\dagger}_j\right)^2  \left(\hat{a}_{k}+\hat{a}^{\dagger}_{k}\right)^2.\nonumber
	\end{eqnarray}
	Using the RWA, we obtain
	\begin{eqnarray}
		&& -\sum_{\langle j,k\rangle}E'_{J,jk}
		\left(\frac{2E_{C,j}}{E^{\rm eq}_{J,j}}\frac{2E_{C,k}}{E^{\rm eq}_{J,k}}\right)^{1/2}
		\left(\hat{n}_j\hat{n}_{k}+\frac{1}{4}\left(\hat{a}^{\dagger}_{j}\right)^2\hat{a}^2_{k}\right.\nonumber\\&&+\left.\frac{1}{4}\hat{a}^2_j\left(\hat{a}^{\dagger}_{k}\right)^2+\frac{1}{2}\hat{n}_j+\frac{1}{2}\hat{n}_{k}\right)=-\sum_{\langle j,k\rangle}\Delta_{jk}
		\left(\hat{n}_j\hat{n}_{k}\right.\nonumber
		\\&&\left.+\frac{1}{4}\left(\hat{a}^{\dagger}_{j}\right)^2\hat{a}^2_{k}+\frac{1}{4}\hat{a}^2_j\left(\hat{a}^{\dagger}_{k}\right)^2+\frac{1}{2}\hat{n}_j+\frac{1}{2}\hat{n}_{k}\right),
	\end{eqnarray}
	where the coupling strengths read
	\begin{eqnarray}
		\Delta_{jk}=2E'_{J,jk}
		\left(\frac{E_{C,j}}{E^{\rm eq}_{J,j}}\frac{E_{C,k}}{E^{\rm eq}_{J,k}}\right)^{1/2}.
	\end{eqnarray}
	We divide the last part of the Hamiltonian into two components. The first component has the form 
	\begin{eqnarray}
		&&\sum_{\langle j,k\rangle} E'_{J,jk}\,\frac{\hat{\phi}^3_{j}\,\hat{\phi}_{k}}{6\,\Phi^4_0}=
		\sum_{\langle j,k\rangle} \frac{E'_{J,jk}}{6}
		\left(\frac{2E_{C,j}}{E^{\rm eq}_{J,j}}\right)^{3/4}
		\left(\frac{2E_{C,k}}{E^{\rm eq}_{J,k}}\right)^{1/4}\nonumber
		\\&&\times\left(\hat{a}^{\dagger}_j+\hat{a}_j\right)^3
		\left(\hat{a}^{\dagger}_{k}+\hat{a}_{k}\right)= 
		\sum_{\langle j,k\rangle} E'_{J,jk}
		\left(\frac{E_{C,j}}{E^{\rm eq}_{J,j}}\right)^{3/4}\nonumber\\&&\times
		\left(\frac{E_{C,k}}{E^{\rm eq}_{J,k}}\right)^{1/4}
		\left(\hat{a}_j\,\hat{n}_j\,\hat{a}^{\dagger}_{k}+\hat{a}_{k}\,\hat{n}_j\,\hat{a}^{\dagger}_{j}\right),
	\end{eqnarray}
	where $\hat{n}_j=\hat{a}^{\dagger}_j\hat{a}_j$ are the number operators. The second component is obtained by interchanging the indices. The resulting expression for the last part of the Hamiltonian reads  
	\begin{eqnarray}
		&&\sum_{\langle j,k\rangle} E'_{J,jk}\,\frac{\hat{\phi}_{j}\,\hat{\phi}_{k}}{6\,\Phi^4_0}\left(\hat{\phi}_{j}^2+\hat{\phi}_{k}^2\right)=
		\sum_{\langle j,k\rangle} E'_{J,jk}
		\left(\frac{E_{C,j}}{E^{\rm eq}_{J,j}}\right)^{3/4}\nonumber\\
		&&\times\left(\frac{E_{C,k}}{E^{\rm eq}_{J,k}}\right)^{1/4}
		\left(\hat{a}_j\,\hat{n}_j\,\hat{a}^{\dagger}_{k}+\hat{a}_{k}\,\hat{n}_j\,\hat{a}^{\dagger}_{j}\right)\nonumber\\
		&&
		+\sum_{\langle j,k\rangle} E'_{J,jk}\left(\frac{E_{C,j}}{E^{\rm eq}_{J,j}}\right)^{1/4}\left(\frac{E_{C,k}}{E^{\rm eq}_{J,k}}\right)^{3/4}
\left(\hat{a}_j\,\hat{n}_{k}\,\hat{a}^{\dagger}_{k}\right.\nonumber\\&&+\left.\hat{a}_{k}\,\hat{n}_{k}\,\hat{a}^{\dagger}_{j}\right)= \sum_{\langle j,k\rangle} T_{jk}
		\left(\hat{a}_j\,\hat{n}_j\,\hat{a}^{\dagger}_{k}+\hat{a}_{k}\,\hat{n}_j\,\hat{a}^{\dagger}_{j}\right)\nonumber\\
		&&+\sum_{\langle j,k\rangle} \overline{T}_{jk}
		\left(\hat{a}_j\,\hat{n}_{k}\,\hat{a}^{\dagger}_{k}+\hat{a}_{k}\,\hat{n}_{k}\,\hat{a}^{\dagger}_{j}\right),
	\end{eqnarray}
	where we define the parameters
	\begin{eqnarray}
		T_{jk} = E'_{J,jk}
		\left(\frac{E_{C,j}}{E^{\rm eq}_{J,j}}\right)^{3/4}
		\left(\frac{E_{C,k}}{E^{\rm eq}_{J,k}}\right)^{1/4},\\
		\overline{T}_{jk}=E'_{J,jk}
		\left(\frac{E_{C,j}}{E^{\rm eq}_{J,j}}\right)^{1/4}
		\left(\frac{E_{C,k}}{E^{\rm eq}_{J,k}}\right)^{3/4}.
	\end{eqnarray}
Combining all the parts, we obtain the 
full Hamiltonian of the circuit 
    $\hat{H}_{\rm JJA}=\hat{H}^{(0)}_{\rm JJA}+\hat{H}^{(1)}_{\rm JJA}$. The first term reads
    \begin{eqnarray}\label{eq:HJJA_0}
		\hat{H}^{(0)}_{\rm JJA}&& = \sum_{j=1}^N\left(\hbar\,\omega_j+\hbar\,\delta\omega_j-\frac{1}{2}\Delta^{\rm eq}_{j}\right)\hat{n}_j
		 \\&&+\sum_{\langle j,k\rangle}t_{jk}\left(\hat{a}_{j}^{\dagger}\hat{a}_{k}+\hat{a}_{j}\hat{a}^{\dagger}_{k}\right)-\sum_{\langle j,k\rangle}\Delta_{jk}\,
		\hat{n}_j\hat{n}_{k}\nonumber\\
		&& +\sum_{\langle j,k\rangle} T_{jk}
\left(\hat{a}_j\,\hat{n}_j\,\hat{a}^{\dagger}_{k}+\hat{a}_{k}\,\hat{n}_j\,\hat{a}^{\dagger}_{j}\right)\nonumber\\
		&&
		+\sum_{\langle j,k\rangle}\overline{T}_{jk}
\left(\hat{a}_j\,\hat{n}_{k}\,\hat{a}^{\dagger}_{k}+\hat{a}_{k}\,\hat{n}_{k}\,\hat{a}^{\dagger}_{j}\right),\nonumber
	\end{eqnarray}
where we introduce the equivalent coupling strengths as 
	\begin{equation}
	   \Delta^{\rm eq}_{j} =  \sum_{k\in \text{NN}(j)}\Delta_{jk}.
	\end{equation}
The superconducting circuit system described by Eq.~(\ref{eq:HJJA_0}) is interpreted as an array of oscillators with frequencies
	$\omega_j$ and nonlinearities $\delta\omega_j$, $\Delta^{\rm eq}_{j}$ interacting via linear hopping parameters 
	$t_{jk}$ and nonlinear cross-Kerr couplings $\Delta_{jk}$, $T_{jk}$ and $\overline{T}_{jk}$.
The second term of the Hamiltonian reads
\begin{eqnarray}
\hat{H}^{(1)}_{\rm JJA} &&= \sum_{j=1}^N
    \frac{\hbar\,\delta\omega_j}{2}(\hat{a}^{\dagger}_j)^2\hat{a}^2_j\\
&&-\sum_{\langle j,k\rangle}\frac{\Delta_{jk}}{4}	\left(\left(\hat{a}^{\dagger}_{j}\right)^2\hat{a}^2_{k}+\hat{a}^2_j\left(\hat{a}^{\dagger}_{k}\right)^2\right).\nonumber
\end{eqnarray} 
It does not effect the dynamics of the JJA provided the system state belongs to the Hilbert space spanned by the physical spin-1/2 states. Such evolution is achieved by the 
initial state condition as well as the proper choice of the parameters of the Hamiltonian $\hat{H}^{(0)}_{\rm JJA}$  so that it is aligned with the EBH model.
    
\section{\label{appendix:parameter_space}Detailed derivation of the parameters of the circuit QED-based simulator}
In this Appendix, we examine the parameter space of the circuit QED-based simulator.  We recast the EBH Hamiltonian (up to a constant) as 
	\begin{eqnarray}\label{eq:EBH_1D}
		\hat{H}_{\rm EBH} &=&-\sum_{\langle j,k\rangle} \frac{J_{jk}}{2}\left(3\,\hat{a}^{\dagger}_j\hat{a}_{k}-\hat{a}_j(\hat{n}_j+\hat{n}_{k})\hat{a}^{\dagger}_{k}+\mbox{h.c.}\right)
		\nonumber\\&&-\sum_{\langle j,k\rangle} J_{jk}\hat{n}_j\hat{n}_{k}+\sum_{j=1}^{N} \hat{n}_j\left(h_j+\frac{1}{2}J^{\rm eq}_j\right),
	\end{eqnarray}
	where we introduce the notation 
	\begin{equation}
        J^{\rm eq}_j =  \sum_{k\in \text{NN}(j)}J_{jk}.
	\end{equation} 
By comparing the EBH (Eq.~(\ref{eq:EBH_1D})) and the JJA (Eq.~(\ref{eq:H_JJA_simplified})) Hamiltonians,  we establish the correspondence between the two models
\begin{eqnarray}
    h_j+\frac{1}{2}J^{\rm eq}_j &=& \hbar\,\omega_j+\hbar\,\delta\omega_j-\frac{1}{2}\Delta^{\rm eq}_j,\label{eq_appendix:h}\\
    -3\,J_{jk} &=& 2\,t_{jk}\label{eq_appendix:Jt}\\
    J_{jk} &=& \Delta_{jk},\label{eq_appendix:JDelta}\\
    J_{jk} &=& 2\,T_{jk} = 2\,\overline{T}_{jk}.\label{eq_appendix:TeqTp}
\end{eqnarray}
Next, we analyze Eqs.~(\ref{eq_appendix:h})--~(\ref{eq_appendix:TeqTp}). We observe that Eq.~(\ref{eq_appendix:TeqTp}) leads to the uniform distribution of the impedance, which is defined by \cite{RamosJJA} $Z_j=\sqrt{L_{\mbox{eq},j}/C_{\mbox{eq},j}}=\left(\Phi_0/e\right)\sqrt{2E_{C,j}/E^{\rm eq}_{J,j}}$, across the lattice. In the following, we omit the index $j$ in the ratio $E_{C,j}/E^{\rm eq}_{J,j}$. As a result, the parameters of the JJA Hamiltonian are connected as follows 
\begin{eqnarray}
    \Delta_{jk} &=& 2\,T_{jk}=2\,\overline{T}_{jk}=2\,E'_{J,jk}\frac{E_{C}}{E^{\rm eq}_{J}},\label{eq_appendix:delta}\\
    t_{jk} &=& E_{\text{coup},jk}\,\left(\frac{E^{\rm eq}_{J}}{2E_{C}}\right)^{1/2}-E'_{J,jk}\,\left(\frac{2E_{C}}{E^{\rm eq}_{J}}\right)^{1/2}.\label{eq_appendix:tj}
\end{eqnarray}
In addition, from Eqs.~(\ref{eq_appendix:Jt}), (\ref{eq_appendix:JDelta}),~(\ref{eq_appendix:delta}), (\ref{eq_appendix:tj}), we obtain  the relation
\begin{eqnarray}\label{eq_appendix:QEDcondition}
		E_{{\rm coup},jk} = 2E'_{J,jk}\frac{E_{C}}{E^{\rm eq}_{J}}
		\left(1-3\left(\frac{E_{C}}{2\,E^{\rm eq}_{J}}\right)^{1/2}\right).
	\end{eqnarray}
Thus, we map the parameters of the Heisenberg model onto the circuit-QED simulator parameters

\begin{eqnarray}
		J_{jk} &=& 2\,E'_{J,jk}\frac{\,E_{C}}{E^{\rm eq}_{J}},\label{eq_appendix:corr_J}\\
		h_j &=&\left(8E_{C,j}\,E^{\rm eq}_{J,j}\right)^{1/2}-3E_{C,j}+
		\frac{2E_{C}}{E^{\rm eq}_{J}}E_{J,j}.\label{eq_appendix:corr_h}
\end{eqnarray}
We are particularly interested in the inverse transformation, which expresses the circuit-QED simulator variables in terms of the coupling constants \( J_{jk} \) and magnetic field strengths \( h_j \). For convenience, we introduce the following notation:
\begin{equation}
    r = \frac{E^{\rm eq}_{J}}{E_C}.\label{eq_appendix:r}
\end{equation}
The  Josephson energies $E'_{J,jk}$ are calculated from Eq.~(\ref{eq_appendix:corr_J})
\begin{equation}
    E'_{J,jk} = \frac{r\,J_{jk}}{2}.
\end{equation}
The capacitive coupling energies are determined by Eq.~\eqref{eq_appendix:QEDcondition}, yielding
\begin{equation}
    E_{{\rm coup},jk} = J_{jk}\left(1-\frac{3}{\sqrt{2\,r}}\right).
\end{equation}
Using Eq.~(\ref{eq:frequency}), we have
\begin{equation}
    r = \frac{\left(\hbar\,\omega_j\right)^2}{8E^2_{C,j}}.
\end{equation}
Thus, the capacitive energies  can be recast as
\begin{equation}
    E_{C,j} = \frac{\hbar\,\omega_j}{\sqrt{8\,r}}.\label{eq_appendix:Ec_r}
\end{equation}
Replacing Eq. (\ref{eq_appendix:Ec_r}) into Eq. (\ref{eq_appendix:h}), we obtain 
\begin{equation}
    h_j = \hbar\,\omega_j - \frac{\hbar\,\omega_j}{\sqrt{8\,r}} -J^{\rm eq}_j.
\end{equation}
Finally, the transformation that determines the circuit-QED simulator parameters is given by  

\begin{eqnarray}
       \hbar\,\omega_j = \frac{h_j+J^{\rm eq}_j}{1-1/\sqrt{8\,r}},\label{eq:inverse_map_1}\\
       \hbar\,\delta\omega_j = -
       \frac{h_j+J^{\rm eq}_j}{\sqrt{8\,r}-1},\label{eq:inverse_map_2}\\
       t_{jk} = -\frac{3}{2}J_{jk},\\
       \Delta_{jk} = J_{jk}, \\
       T_{jk} = \overline{T}_{jk} = \frac{J_{jk}}{2}.
\end{eqnarray}
Equations \eqref{eq:inverse_map_1} and \eqref{eq:inverse_map_2} depend on the parameter $r$. This implies that the same set of spin couplings $J_{jk}$ and magnetic fields $h_j$ can be realized by multiple configurations of the simulator parameters.

\newpage
\bibliography{bibliography}
\newpage
\end{document}